\documentclass[adraft]{eptcs}
\pdfoutput=1
\usepackage{underscore}

\usepackage{amsmath,amsthm}
\usepackage{amssymb}

\usepackage[capitalize,noabbrev,nameinlink]{cleveref} 

\usepackage[numbers]{natbib}

\usepackage{stmaryrd} 

\usepackage{wasysym} 

\usepackage{mathtools}

\newcommand\xlrsquigarrow{%
  \mathrel{%
    \vcenter{%
      \hbox{%
        \begin{tikzpicture}
          \path[
            draw,
            >={Implies[]},
            <->,
            double distance between line centers=1.5pt,
            decorate,
            decoration={
              zigzag,
              amplitude=0.7pt,
              segment length=3pt,
              pre length=4pt,
              post length=4pt,
            },
          ]   
            (0,0) -- (14pt,0);
        \end{tikzpicture}%
      }%
    }%
  }%
}

\usepackage[utf8]{inputenc}

\usepackage{newunicodechar}
\let\Alpha=A
\let\Beta=B
\let\Epsilon=E
\let\Zeta=Z
\let\Eta=H
\let\Iota=I
\let\Kappa=K
\let\Mu=M
\let\Nu=N
\let\Omicron=O
\let\omicron=o
\let\Rho=P
\let\Tau=T
\let\Chi=X

\newunicodechar{Α}{\ensuremath{\Alpha}}
\newunicodechar{α}{\ensuremath{\alpha}}
\newunicodechar{Β}{\ensuremath{\Beta}}
\newunicodechar{β}{\ensuremath{\beta}}
\newunicodechar{Γ}{\ensuremath{\Gamma}}
\newunicodechar{γ}{\ensuremath{\gamma}}
\newunicodechar{Δ}{\ensuremath{\Delta}}
\newunicodechar{δ}{\ensuremath{\delta}}
\newunicodechar{Ε}{\ensuremath{\Epsilon}}
\newunicodechar{ε}{\ensuremath{\epsilon}}
\newunicodechar{ϵ}{\ensuremath{\varepsilon}}
\newunicodechar{Ζ}{\ensuremath{\Zeta}}
\newunicodechar{ζ}{\ensuremath{\zeta}}
\newunicodechar{Η}{\ensuremath{\Eta}}
\newunicodechar{η}{\ensuremath{\eta}}
\newunicodechar{Θ}{\ensuremath{\Theta}}
\newunicodechar{θ}{\ensuremath{\theta}}
\newunicodechar{ϑ}{\ensuremath{\vartheta}}
\newunicodechar{Ι}{\ensuremath{\Iota}}
\newunicodechar{ι}{\ensuremath{\iota}}
\newunicodechar{Κ}{\ensuremath{\Kappa}}
\newunicodechar{κ}{\ensuremath{\kappa}}
\newunicodechar{Λ}{\ensuremath{\Lambda}}
\newunicodechar{λ}{\ensuremath{\lambda}}
\newunicodechar{Μ}{\ensuremath{\Mu}}
\newunicodechar{μ}{\ensuremath{\mu}}
\newunicodechar{Ν}{\ensuremath{\Nu}}
\newunicodechar{ν}{\ensuremath{\nu}}
\newunicodechar{Ξ}{\ensuremath{\Xi}}
\newunicodechar{ξ}{\ensuremath{\xi}}
\newunicodechar{Ο}{\ensuremath{\Omicron}}
\newunicodechar{ο}{\ensuremath{\omicron}}
\newunicodechar{Π}{\ensuremath{\Pi}}
\newunicodechar{π}{\ensuremath{\pi}}
\newunicodechar{ϖ}{\ensuremath{\varpi}}
\newunicodechar{Ρ}{\ensuremath{\Rho}}
\newunicodechar{ρ}{\ensuremath{\rho}}
\newunicodechar{ϱ}{\ensuremath{\varrho}}
\newunicodechar{Σ}{\ensuremath{\Sigma}}
\newunicodechar{σ}{\ensuremath{\sigma}}
\newunicodechar{ς}{\ensuremath{\varsigma}}
\newunicodechar{Τ}{\ensuremath{\Tau}}
\newunicodechar{τ}{\ensuremath{\tau}}
\newunicodechar{Υ}{\ensuremath{\Upsilon}}
\newunicodechar{υ}{\ensuremath{\upsilon}}
\newunicodechar{Φ}{\ensuremath{\Phi}}
\newunicodechar{φ}{\ensuremath{\varphi}}
\newunicodechar{ϕ}{\ensuremath{\phi}}
\newunicodechar{Χ}{\ensuremath{\Chi}}
\newunicodechar{χ}{\ensuremath{\chi}}
\newunicodechar{Ψ}{\ensuremath{\Psi}}
\newunicodechar{ψ}{\ensuremath{\psi}}
\newunicodechar{Ω}{\ensuremath{\Omega}}
\newunicodechar{ω}{\ensuremath{\omega}}

\newunicodechar{ℂ}{\ensuremath{\mathbb{C}}}
\newunicodechar{ℤ}{\ensuremath{\mathbb{Z}}}
\newunicodechar{ℝ}{\ensuremath{\mathbb{R}}}
\newunicodechar{ℕ}{\ensuremath{\mathbb{N}}}
\newunicodechar{ℚ}{\ensuremath{\mathbb{Q}}}

\newunicodechar{𝒜}{\ensuremath{\mathcal{A}}}
\newunicodechar{ℬ}{\ensuremath{\mathcal{B}}}
\newunicodechar{𝒞}{\ensuremath{\mathcal{C}}}
\newunicodechar{𝒟}{\ensuremath{\mathcal{D}}}
\newunicodechar{ℰ}{\ensuremath{\mathcal{E}}}
\newunicodechar{ℱ}{\ensuremath{\mathcal{F}}}
\newunicodechar{𝒢}{\ensuremath{\mathcal{G}}}
\newunicodechar{ℋ}{\ensuremath{\mathcal{H}}}
\newunicodechar{ℑ}{\ensuremath{\mathcal{I}}}
\newunicodechar{𝒦}{\ensuremath{\mathcal{K}}}
\newunicodechar{ℒ}{\ensuremath{\mathcal{L}}}
\newunicodechar{ℳ}{\ensuremath{\mathcal{M}}}
\newunicodechar{𝒩}{\ensuremath{\mathcal{N}}}
\newunicodechar{𝒪}{\ensuremath{\mathcal{O}}}
\newunicodechar{𝒫}{\ensuremath{\mathcal{P}}}
\newunicodechar{𝒬}{\ensuremath{\mathcal{Q}}}
\newunicodechar{ℛ}{\ensuremath{\mathcal{R}}}
\newunicodechar{ℜ}{\ensuremath{\mathcal{R}}}
\newunicodechar{𝒮}{\ensuremath{\mathcal{S}}}
\newunicodechar{𝒯}{\ensuremath{\mathcal{T}}}
\newunicodechar{𝒰}{\ensuremath{\mathcal{U}}}
\newunicodechar{𝒱}{\ensuremath{\mathcal{V}}}
\newunicodechar{𝒲}{\ensuremath{\mathcal{W}}}
\newunicodechar{ℓ}{\ensuremath{\ell}}

\newunicodechar{∼}{\ensuremath{\sim}}
\newunicodechar{≈}{\ensuremath{\mathrel{\approx}}}
\newunicodechar{≋}{\ensuremath{\bumpeq}}
\newunicodechar{≅}{\ensuremath{\cong}}
\newunicodechar{≡}{\ensuremath{\equiv}}
\newunicodechar{≂}{\ensuremath{\mathrel{\aquarius}}}
\newunicodechar{≤}{\ensuremath{\le}}
\newunicodechar{≥}{\ensuremath{\ge}}
\newunicodechar{≲}{\ensuremath{\lesssim}}
\newunicodechar{≠}{\ensuremath{\neq}}
\newunicodechar{≔}{\ensuremath{\coloneqq}}
\newunicodechar{⋆}{\ensuremath{\applysymbol}}
\newunicodechar{ƛ}{\ensuremath{\linearlambda}}

\newunicodechar{±}{\ensuremath{\pm}}
\newunicodechar{∓}{\ensuremath{\pm}}
\newunicodechar{×}{\ensuremath{\times}}
\newunicodechar{×}{\times}

\newunicodechar{→}{\ensuremath{\rightarrow}}
\newunicodechar{←}{\ensuremath{\leftarrow}}
\newunicodechar{⇒}{\ensuremath{\Rightarrow}}
\newunicodechar{↦}{\ensuremath{\mapsto}}
\newunicodechar{↝}{\ensuremath{\leadsto}}
\newunicodechar{⸖}{\ensuremath{\mathrel{>!}}}
\newunicodechar{⇆}{\ensuremath{\xlrsquigarrow}}
\newunicodechar{↔}{\ensuremath{\leftrightarrow}}
\newunicodechar{↭}{\ensuremath{\leftrightsquigarrow}}
\newunicodechar{↣}{\ensuremath{\rightarrowtail}}

\newunicodechar{∀}{\ensuremath{\forall}}
\newunicodechar{∃}{\ensuremath{\exists}}
\newunicodechar{∨}{\ensuremath{\vee}}
\newunicodechar{∧}{\ensuremath{\wedge}}
\newunicodechar{⊢}{\ensuremath{\vdash}}
\newunicodechar{⊣}{\ensuremath{\dashv}}
\newunicodechar{⊤}{\ensuremath{\top}}
\newunicodechar{⊥}{\ensuremath{\bot}}
\newunicodechar{¬}{\ensuremath{\neg}}

\newunicodechar{⊸}{\ensuremath{\multimap}}
\newunicodechar{⊗}{\ensuremath{\otimes}}
\newunicodechar{⊕}{\ensuremath{\oplus}}
\newunicodechar{¡}{\ensuremath{\oc}}
\newunicodechar{⨂}{\ensuremath{\bigotimes}}
\newunicodechar{⨁}{\ensuremath{\bigoplus}}

\newunicodechar{∈}{\ensuremath{\in}}
\newunicodechar{∉}{\ensuremath{\not\in}}
\newunicodechar{⊆}{\ensuremath{\subseteq}}
\newunicodechar{∪}{\ensuremath{\cup}}
\newunicodechar{⋓}{\ensuremath{\Cup}}
\newunicodechar{∅}{\ensuremath{\emptyset}}

\newunicodechar{〈}{\ensuremath{\langle}}
\newunicodechar{⟨}{\ensuremath{\langle}}
\newunicodechar{⟩}{\ensuremath{\rangle}}
\newunicodechar{〉}{\ensuremath{\rangle}}
\newunicodechar{⟦}{\ensuremath{\llbracket}}
\newunicodechar{⟧}{\ensuremath{\rrbracket}}

\newunicodechar{₀}{\ensuremath{_0}}
\newunicodechar{₁}{\ensuremath{_1}}
\newunicodechar{₂}{\ensuremath{_2}}
\newunicodechar{₃}{\ensuremath{_3}}
\newunicodechar{₄}{\ensuremath{_4}}
\newunicodechar{₅}{\ensuremath{_5}}
\newunicodechar{₆}{\ensuremath{_6}}
\newunicodechar{₇}{\ensuremath{_7}}
\newunicodechar{₈}{\ensuremath{_8}}
\newunicodechar{₉}{\ensuremath{_9}}
\newunicodechar{ₐ}{\ensuremath{_a}}
\newunicodechar{ₑ}{\ensuremath{_e}}
\newunicodechar{ₕ}{\ensuremath{_h}}
\newunicodechar{ᵢ}{\ensuremath{_i}}
\newunicodechar{ⱼ}{\ensuremath{_j}}
\newunicodechar{ₖ}{\ensuremath{_k}}
\newunicodechar{ₗ}{\ensuremath{_l}}
\newunicodechar{ₘ}{\ensuremath{_m}}
\newunicodechar{ₙ}{\ensuremath{_n}}
\newunicodechar{ₒ}{\ensuremath{_o}}
\newunicodechar{ₚ}{\ensuremath{_p}}
\newunicodechar{ᵣ}{\ensuremath{_r}}
\newunicodechar{ₛ}{\ensuremath{_s}}
\newunicodechar{ₜ}{\ensuremath{_t}}
\newunicodechar{ᵤ}{\ensuremath{_u}}
\newunicodechar{ₓ}{\ensuremath{_x}}

\newunicodechar{⁰}{\ensuremath{^0}}
\newunicodechar{¹}{\ensuremath{^1}}
\newunicodechar{²}{\ensuremath{^2}}
\newunicodechar{³}{\ensuremath{^3}}
\newunicodechar{⁴}{\ensuremath{^4}}
\newunicodechar{⁵}{\ensuremath{^5}}
\newunicodechar{⁶}{\ensuremath{^6}}
\newunicodechar{⁷}{\ensuremath{^7}}
\newunicodechar{⁸}{\ensuremath{^8}}
\newunicodechar{⁹}{\ensuremath{^9}}
\newunicodechar{ᵃ}{\ensuremath{^a}}
\newunicodechar{ᵇ}{\ensuremath{^b}}
\newunicodechar{ᶜ}{\ensuremath{^c}}
\newunicodechar{ᵈ}{\ensuremath{^d}}
\newunicodechar{ᵉ}{\ensuremath{^e}}
\newunicodechar{ᶠ}{\ensuremath{^f}}
\newunicodechar{ᵍ}{\ensuremath{^g}}
\newunicodechar{ʰ}{\ensuremath{^h}}
\newunicodechar{ⁱ}{\ensuremath{^i}}
\newunicodechar{ʲ}{\ensuremath{^j}}
\newunicodechar{ᵏ}{\ensuremath{^k}}
\newunicodechar{ᵐ}{\ensuremath{^m}}
\newunicodechar{ⁿ}{\ensuremath{^n}}
\newunicodechar{ᵒ}{\ensuremath{^o}}
\newunicodechar{ᵖ}{\ensuremath{^p}}
\newunicodechar{ˢ}{\ensuremath{^s}}
\newunicodechar{ᵗ}{\ensuremath{^t}}
\newunicodechar{ᵘ}{\ensuremath{^u}}
\newunicodechar{ʷ}{\ensuremath{^v}}
\newunicodechar{ˣ}{\ensuremath{^x}}
\newunicodechar{ʸ}{\ensuremath{^y}}
\newunicodechar{ᶻ}{\ensuremath{^z}}

\newunicodechar{•}{\ensuremath{\bullet}}
\newunicodechar{∙}{\ensuremath{\bullet}}
\newunicodechar{·}{\ensuremath{\cdot}}
\newunicodechar{⋯}{\ensuremath{\cdots}}
\newunicodechar{…}{\ensuremath{\ldots}}
\newunicodechar{∷}{\ensuremath{~\mathrel{:\!\!\!:}~}}

\newunicodechar{∣}{\ensuremath{\mid}}
\newunicodechar{∥}{\ensuremath{ {\parallel} }}

\newunicodechar{□}{\ensuremath{\square}}

\newunicodechar{∗}{\ensuremath{\ast}}

\newunicodechar{∘}{\ensuremath{\circ}}
\newunicodechar{†}{\ensuremath{^{\dagger}}}
\newunicodechar{♯}{\ensuremath{\mathrel{\#}}}

\newunicodechar{∞}{\ensuremath{\infty}}
\newunicodechar{£}{\ensuremath{\mathrel{\$}}}

\newunicodechar{⧺}{\ensuremath{\doubleplus}}

\usepackage{aliascnt}

\crefformat{equation}{#2Equation~#1#3}
\crefname{appendix}{Appendix}{Appendices}

\usepackage{thmtools,thm-restate}

\declaretheorem[name=Theorem
               ,refname={Theorem,Theorems}
               ,Refname={Theorem,Theorems}]
               {theorem}
\newtheorem{lemma}[theorem]{Lemma}
\newtheorem{axiom}[theorem]{Axiom}
\newtheorem{definition}[theorem]{Definition}
\newtheorem{proposition}[theorem]{Proposition}

\usepackage{url}

\usepackage{breakurl}

\usepackage{mathpartir}

\usepackage{booktabs}

\usepackage{xcolor}
\definecolor{ltblue}{rgb}{0,0.4,0.4}
\definecolor{dkblue}{rgb}{0,0.1,0.6}
\definecolor{dkgreen}{rgb}{0,0.35,0}
\definecolor{dkviolet}{rgb}{0.3,0,0.5}
\definecolor{dkred}{rgb}{0.5,0,0}

\usepackage{listings}
\input{lstcoq.sty}
\input{lsthaskell.sty}

\usepackage{tikz} 
\usetikzlibrary{cd}
\usetikzlibrary{positioning}
\usetikzlibrary{%
  arrows,%
  shapes.misc,
  shapes.arrows,%
  shapes.callouts,
  chains,%
  matrix,%
  positioning,
  scopes,%
  decorations.pathmorphing,
  decorations.text,
  shadows%
}

\newcommand{\kwfont}[1]{\textup{\texttt{#1}}} 
\usepackage{textcomp} 

\usepackage{xparse}
\NewDocumentCommand{\optionalParens}{s m m}{
    \IfBooleanTF{#2}{(#3)}{\IfBooleanTF{#1}{~#3}{#3}}
}
\NewDocumentCommand{\apply}{m s O{} m}{
    #1#3 \optionalParens*{#2}{#4}
}

\NewDocumentCommand{\applytwo}{m O{} s m s m}{ 
   #1#2~\optionalParens{#3}{#4}~\optionalParens{#5}{#6}
}

\newcommand{\tagsc}[1]{\tag{\textsc{#1}}}

\usepackage{xspace}

\newcommand{\ie}{\emph{i.e.,}\xspace}

\usepackage{cmll} 

\usepackage{braket}

\def\Bool{\kwfont{Bool}}

\def\Qubit{\kwfont{Qubit}}

\def\Type{\kwfont{Type}}
\def\FinType{\kwfont{FinType}}
\def\LType{\kwfont{QType}}
\newcommand{\LTypeVar}{\kwfont{TVar}}
\def\QType{\kwfont{QType}}

\def\QEXP{\kwfont{QExp}}
\def\BEXP{\kwfont{BExp}}

\def\Ctx{\kwfont{Ctx}}

\def\DENSITY{\kwfont{Density}}

\def\CStar{C^\ast}

\def\CStarCPU{C^\ast_{\text{CPU}}}
\def\Density{\apply\DENSITY}

\def\Vec{\applytwo{\kwfont{Vec}}}

\newcommand{\LExp}{\applytwo{\QEXP}}
\newcommand{\BExp}{\applytwo{\BEXP}}

\newcommand{\Var}{\kwfont{Var}}
\newcommand{\BVar}{\kwfont{\underline{Var}}}

\newcommand{\Void}{\kwfont{Void}}

\newcommand{\LZERO}{\kwfont{LZERO}}

\def\LET{\kwfont{let}}
\NewDocumentCommand{\letin}{msmm}{
  \LET~#1 ≔ \optionalParens{#2}{#3}~\kwfont{in}~#4
}
\NewDocumentCommand{\caseof}{s m m m}{
    \kwfont{case}\IfBooleanTF{#1}{(#2)}{~#2}~\kwfont{of}~(#3 \mid #4)
}

\newcommand{\casesof}[3]{
    \kwfont{case}~#1~\kwfont{of}
    \begin{cases}
      #2 \\
      #3
    \end{cases}
}
\newcommand{\emptycase}[1]{\kwfont{case}~#1~\kwfont{of}~()}

\NewDocumentCommand{\matchwith}{m s m m}{
    \kwfont{match}_{#1}\IfBooleanTF{#2}{(#3)}{~#3}~\kwfont{with}~#4
}
\newcommand{\MATCH}{\kwfont{match}}
\newcommand{\matchRender}[3]{\MATCH_{#1}~#2~\kwfont{with}~#3}
\NewDocumentCommand{\match}{s m s m s m}{
    \matchRender{\IfBooleanTF{#1}{(#2)}{#2}}
                {\IfBooleanTF{#3}{(#4)}{#4}}
                {\IfBooleanTF{#5}{(#6)}{#6}}
}

\newcommand{\IF}{\kwfont{if}}
\newcommand{\THEN}{\kwfont{then}}
\newcommand{\ELSE}{\kwfont{else}}
\newcommand{\IfThen}[2]{\IF~#1~\THEN~#2}

\newcommand{\IfThenElse}[3]{\IfThen{#1}{#2}~\ELSE~#3}
\newcommand{\inl}{\kwfont{inl}}
\newcommand{\inr}{\kwfont{inr}}
\newcommand{\inj}[1]{\kwfont{in}_{#1}}

\newcommand{\true}{\kwfont{true}}
\newcommand{\false}{\kwfont{false}}

\newcommand{\discard}{\kwfont{discard}}
\newcommand{\discardin}[2]{\discard~#1~\kwfont{in}~#2}

\newcommand{\applysymbol}{\ensuremath{\mathrel{\hat{}}}}
\newcommand{\linearlambda}{\ensuremath{\hat{\lambda}}}

\newcommand{\LOWER}{\kwfont{Lower}}

\newcommand{\Lower}{\apply{\LOWER}}

\newcommand{\PUT}{\kwfont{put}}
\newcommand{\SUSPEND}{\kwfont{suspend}}
\newcommand{\FORCE}{\kwfont{force}}
\renewcommand{\put}{\apply{\PUT}}
\newcommand{\suspend}{\apply{\SUSPEND}}
\newcommand{\force}{\apply{\FORCE}}

\newcommand{\lbind}{\ensuremath{\mathbin{=\!\!>\!\!>\!\!=}}}

\newcommand{\unitary}[2]{{#1} ♯ {#2}}

\newcommand{\CNOT}{\kwfont{CNOT}}
\newcommand{\cnot}{\kwfont{cnot}}

\newcommand{\qwire}{\ensuremath{\mathcal{Q}\textsc{wire}}\xspace}

\newcommand{\liquid}{LIQ$Ui|\rangle$\xspace}

\newcommand{\MEAS}{\kwfont{meas}}
\newcommand{\meas}{\apply{\MEAS}}
\newcommand{\INIT}{\kwfont{init}}
\newcommand{\init}{\apply{\INIT}}
\newcommand{\NEW}{\kwfont{new}}
\newcommand{\new}{\apply{\NEW}}

\NewDocumentCommand{\boxQ}{s m m}{
  \IfBooleanTF{#1}{\kwfont{box\_}~#2 ⇒ #3}
                  {\kwfont{box}~#2 ⇒ #3}
}

\usepackage{array}
\usepackage{multicol}

\newcommand{\LUNIT}{\kwfont{lunit}}

\newcommand{\ID}{\kwfont{id}}

\newcommand{\ASSOC}{\kwfont{ASSOC}}
\NewDocumentCommand{\assoc}{ggg}{
  \IfNoValueTF{#1}
   {\kwfont{assoc}}
   {\kwfont{assoc}_{#1,#2,#3}}
}

\newcommand{\SWAP}{\kwfont{SWAP}}
\NewDocumentCommand{\swap}{oo}{
    \IfNoValueTF{#1}
    {\kwfont{swap}} 
    {\kwfont{swap}_{#1,#2}} 
}

\newcommand{\distr}{\kwfont{distr}}
\newcommand{\DISTR}{\kwfont{DISTR}}

\newcommand{\UNIV}{\kwfont{univ}}
\newcommand{\univ}{\apply{\UNIV}}
\newcommand{\ap}{\apply{\kwfont{ap}}}
\newcommand{\apTwo}{\applytwo{\kwfont{ap}^2}}
\newcommand{\apd}[1]{\kwfont{apd}_{#1}}
\newcommand{\TRANSPORT}{\kwfont{transport}}
\newcommand{\transport}[3]{\TRANSPORT_{#1}~#2~#3}

\newcommand{\coerce}[2]{\kwfont{coerce}~#1~#2}
\newcommand{\pathind}[1]{\kwfont{path\_ind}_{#1}}
\newcommand{\REC}{\kwfont{rec}}
\newcommand{\IND}{\kwfont{ind}}
\newcommand{\rec}{\apply{\REC}}

\newcommand{\quotient}[2]{#1 /_1 #2}
\newcommand{\POINT}{\kwfont{point}}
\newcommand{\point}{\apply{\POINT}}
\newcommand{\CELL}{\kwfont{cell}}
\newcommand{\cell}{\apply\CELL}
\newcommand{\CELLCOMPOSE}{\kwfont{cell\_compose}}
\newcommand{\cellcompose}[2]{\CELLCOMPOSE~#1~#2}
\newcommand{\ONETYPE}{\kwfont{1-type}}
\newcommand{\ISONETYPE}{\kwfont{is-1-type}}
\newcommand{\onetype}{\apply{\ONETYPE}}

\newcommand{\PCell}[1]{\kwfont{P\_cell}_{#1}}
\newcommand{\PPoint}[1]{\kwfont{P\_point}_{#1}}

\newcommand{\Matrix}{\kwfont{Matrix}}
\newcommand{\UMatrix}{\kwfont{UMatrix}}

\newcommand{\fromS}[1]{⟨#1⟩}
\NewDocumentCommand{\toS}{O{} m}{
    \llparenthesis #2 \rrparenthesis^{#1}
}

\usepackage{comment}
\usepackage{draft}
\drafttrue

\newnote[JP]{jennifer}{red}
\newnote[SZ]{steve}{blue}
\newnote{fixme}{green}
\newnote{todo}{green}
\newnote[Cite:]{tocite}{green}

\makeatletter
\NewStdDefinerWrappers{DraftFormat}{%
 \ifdraft
   \expandafter\@firstoftwo
 \else
   \expandafter\@secondoftwo
 \fi
 {#5}{\@firstofone}%
}

\usepackage[normalem]{ulem}
\newcommand*{\coloruwave}[1]{%
 \bgroup
 \markoverwith{\lower3.5\p@\hbox{\sixly\textcolor{#1}{\char58}}}%
 \ULon}
\NewDraftFormat{\awk}{\coloruwave{red}}
\makeatother

\usepackage{enumerate}

    \drafttrue

    \title{A HoTT Quantum Equational Theory (Extended Version)}
    
    \author{Jennifer Paykin
        \institute{Galois Inc. \\ Portland, OR, USA}
        \email{jpaykin@gmail.com}
    \and
        Steve Zdancewic
        \institute{University of Pennsylvania \\ Philadelphia, PA, USA}
        \email{stevez@cis.upenn.edu}
    }
    \def\authorrunning{J. Paykin \& S. Zdancewic}
    \def\titlerunning{HoTT Quantum Equational Theory}

\begin{document}

\maketitle

\begin{abstract}
   This paper presents an equational theory for the QRAM model of quantum
computation, formulated as an embedded language inside of homotopy type theory.
The embedded language approach is highly expressive, and reflects the style of
state-of-the art quantum languages like Quipper and QWIRE. The embedding takes
advantage of features of homotopy type theory to encode unitary transformations
as higher inductive paths, simplifying the presentation of an equational theory.
We prove that this equational theory is sound and complete with respect
to established models of quantum computation.


\end{abstract}

\maketitle

\section{Introduction}
\label{sec:intro}
One of the most prominent models of quantum computation today is known as the
QRAM model, in which a quantum computer works alongside a classical computer to
manipulate both quantum and classical data~\citep{knill1996}. Programming
languages for the QRAM model must provide good access to both quantum data, like
qubits, as well as easy-to-use classical abstractions such as booleans,
recursion, and functions. In addition, the two sorts of data should
interact---it should be possible to measure a qubit and probabilistically obtain
a classical boolean value.

Inspired by the QRAM model, several state-of-the-art quantum programming
languages are implemented as embedded languages, where domain-specific features
for quantum data are added to an existing general-purpose host language. Quipper, embedded in Haskell, utilizes Haskell
type classes, monadic programming, and meta-programming~\citep{green2013}.
\liquid, embedded in F\#, and Chisel-Q, embedded in Scala, also take advantage
of features in their respective host
languages~\citep{wecker2014,liu2013}. Python boasts a large collection of APIs
for quantum computing, including QuTiP~\citep{johansson2013},
QISKit,\footnote{https://github.com/QISKit} and
pyQuil.\footnote{http://pyquil.readthedocs.io}

Unfortunately, reasoning about embedded quantum languages has largely been seen
as futile; such reasoning would have to account for the quantum behavior of the
embedded language as well as the classical behavior of the host language and the
interactions between the two. Attempts to formalize Quipper have been restricted
to a standalone language in the style of Quipper, not the actual
implementation in Haskell~\citep{ross2015,rios2017,mahmoud2018}.

Given a powerful enough host language, however, it is possible to study the
meta-theory of an embedded language inside the host language itself. This is the
approach is taken by \qwire, a language implemented in Coq that uses its host
language both for programming and for verification~\citep{paykin2017,rand2017}.

This work extends that technique to study the equational theory of an embedded
quantum language. We build on an equational theory proposed by
Staton~\cite{staton2015}, an algebraic account of the relationship between quantum
data and classical control. Other works axiomatize particular sets of unitary
transformations~\citep{amy2017,matsumoto2008,nam2017}, but, like
Staton, we focus on the relationships between quantum and
classical features. Staton's algebraic framework includes measurement-based
branching, but it does not contain explicit classical data or features we would
expect from a QRAM-style language.

Because our goal is to extend Staton's equational theory to a richer embedded
language, we choose a host language that specializes in equality---homotopy type
theory (HoTT). In HoTT, proofs of equality, also called \textit{paths}, can
contain extra computational content. In the past few years, a variety of
applications have used paths in HoTT to represent data structures such as
containers~\citep{abbott2004}, version control patches~\citep{angiuli2014}, and
SQL queries~\citep{chu2017}. These data structures can all be represented as groupoids, which are generalized by paths in HoTT.

This work builds on the observation that unitary transformations, a core
component of quantum computing, form a groupoid. We exploit this structure to
encode unitaries in the paths between quantum types. This simplifies Staton's
theory because many of the structural rules in his presentation can be derived
in our presentation.

This paper makes the following contributions:
\begin{itemize}

\item We present a quantum programming language embedded in homotopy type theory
  (\cref{sec:lambda,sec:types}). The language is both (1) expressive, because
  users have access to classical higher-order functions, data structures, and
  more through the host language; and (2) sound, thanks to its use of linear
  types for quantum data. We have formalized some basic properties of the underlying HoTT data structures in Coq.

\item We define an equational theory for the embedded quantum language,
  consisting of standard rules of $α$ and $β$ equivalence, as well as variations
  of $η$ equivalence and commuting conversion rules. In addition, the
  equational theory contains two quantum-specific axioms, which describe how
  unitary transformations interact with initialization and measurement,
  respectively~(\cref{sec:axioms}).

\item We prove the equational theory is sound with respect to a standard
  semantics in terms of superoperators over density matrices
  (\cref{sec:semantics}).

\item Throughout, we use Staton's algebraic equational
  theory~(\cref{sec:statonTheory}) as a specification of the equations we expect
  to hold in our theory, and we prove that a fragment of our language is
  sound and complete with respect to Staton's axioms~(\cref{sec:completeness}).

\end{itemize}

A prior version of this work has been presented in Chapters~6 and~7 of the first author's dissertation~\citep{paykin2018}. The previous version did not include the formal results in Coq, nor the connection we make in \cref{sec:completeness} with Staton's algebraic theory.


\section{Background and main ideas}
\label{sec:background}
\subsection{Quantum computing}

Quantum computers present a radically different computing environment compared
to ordinary classical computers.\footnote{A thorough introduction to quantum
  computation is beyond the scope of this work; we refer the interested
  reader to Nielsen and Chuang~\cite{nielsen2010}.} Instead of bits, quantum computers operate on
\emph{qubits}---superpositions of classical bits of the form
$c₀ \ket{0} + c₁ \ket{1}$ where $c₀,c₁ : ℂ$ are amplitudes satisfying
$|c₀|^2 + |c₁|^2 = 1$. A qubit can be measured, resulting in the bit $0$ with
probability $|c₀|^2$, or the bit $1$ with probability $|c₁|^2$. Qubits~$e$ can
also be manipulated by applying one of a set of \emph{unitary} matrices $U$,
with application written $U ♯ e$. For example, the $X$ (pronounced ``not'')
unitary swaps the amplitudes of $\ket{0}$ and $\ket{1}$, so measuring a qubit of
the form $X ♯ e$ is the same as (results in the same probability distribution as) negating the measurement of~$e$:
$ \meas*{X ♯ e} = ¬ (\meas{e})$.
 
Quantum programs, therefore, do four main things: they initialize qubits, apply
unitary gates, measure qubits, and invoke classical programs to process
classical measurement results. This means that quantum programs are
low-level---they often lack basic abstractions like data structures and
parametricity---and effectful---measuring a qubit can non-locally affect a
different part of the quantum state. 

Researchers realized early on that quantum programming
languages would benefit from formal study, including type
systems~\citep{selinger2009,altenkirch2009},
verification~\citep{ying2012a,rand2017}, and denotational
semantics~\citep{abramsky2008}.
For example, this work uses linear types to enforce the \emph{no-cloning}
principle of quantum mechanics: unknown quantum states cannot be duplicated.
This means that a program that uses quantum data twice, like $λ x. (x,x)$, might
not correspond to a valid quantum computation. Linear types enforce the fact
that quantum variables occur exactly once in a term, so that well-typed programs
have a sound denotational semantics~\citep{selinger2009,selinger2008}

This work also focuses on equational theories, which characterize when two
quantum programs are equivalent. Equational theories help programmers understand
the meaning of their programs and help compiler writers ensure that their
optimizations are sound.
%
%
Unfortunately, developing sound equational theories for effectful languages in general, and quantum languages in particular, is
notoriously difficult.

\subsection{Homotopy type theory (HoTT)}

Homotopy type theory is, in many ways, a theory of equivalence. In HoTT,
proofs of equality $a = b$, called \emph{paths}, may have computational
content. That is, there may be other proofs of equality besides
the trivial reflexivity path $1_a : a = a$.

In homotopy type theory, we write the type of \emph{propositional} proofs of
equality as $a = b$; that is, $a = b$ is a type with a single constructor
$1_a : a = a$. Propositional equality is distinguished from \emph{judgmental}
equality $a ≡ b$, which asserts that $a$ and $b$ are equal by definition. The
judgment $a ≡ b$ is not a type; it is only valid in the meta-theory and has no
computational content. For more intuition on the difference between
propositional and judgmental equality, see the HoTT book~\citep[Chapter
1]{hott2013}.

Homotopy type theory was developed as a type-theoretic alternative to set
theory, but it has applications in a wide variety of computational
domains~\citep{angiuli2014,chu2017,abbott2004}.
When a domain is difficult to characterize equationally, but uses data in the
shape of an equivalence relation or groupoid, HoTT can help.

Consider a type $A$ that you want to quotient by (the equivalence relation generated by) a relation $R$. For every
element $a : A$, the equivalence class of $a$ is written $[a]_R : A/R$, and
whenever $R(a,b)$, it should be the case that $[a]_R = [b]_R$. In set theory it
is possible to define the equivalence class as a set
$[a]_R = \{x : A ∣ R(a,x)\}$, so that $[a]_R$ contains the same elements as
$[b]_R$. However, in programming environments, where the representation of data
structures matters, sets are often implemented as lists or arrays, so $[a]_R$
must be carefully constructed so that it has the same representation as $[b]_R$.

Homotopy type theory sidesteps this representation problem with \emph{higher
  inductive types}: inductive definitions with constructors for both terms and
paths.\footnote{\cref{defn:setQuotient} is an informal, intuitive definition; see Sojakova~\cite{sojakova2015} for a formal definition of set quotients with types and induction principles specified. In that work, Sojakova formalizes higher inductive types as homotopy-intial algebras.}
\begin{definition}[Informal] \label{defn:setQuotient}
  The quotient $A/R$ of a type $A$ by an equivalence relation $R$ on $A$ is
  a higher inductive type generated by the following constructors:
  \begin{enumerate}[-]
    \item for $a : A$, there is a term $[a]_R : A/R$; and
    \item for $a,b : A$ and $r : R(a,b)$, there is a proof $[r]_R:[a]_R=[b]_R$.
  \end{enumerate}
\end{definition}

Notice that if $r₁$ and $r₂$ are two different witnesses (proofs) of $R(a,b)$,
then $[r₁]_R$ is different from $[r₂]_R$---the structure of the relation $R$ is
preserved in the paths of $A/R$.

Despite the extra computational content of equality types in HoTT, the usual
properties still hold for paths generated by higher inductive types.
The principle of \emph{path induction} states that, given a property
$P : \prod_{a,b : α} a = b → \Type$ on paths, if $P$ holds on the reflexivity path,
then $P$ holds on any path. That is, the induction principle for paths
has the following type:
  \begin{align*}
    \pathind{P} &: \left(\prod (x : α), P_{x,x} (1_x)\right)
                → \prod (x~y : α) (p : x = y),~ P_{x,y}(p).
  \end{align*}

  If $p : a = b$ for $a,b: α$ and $x : Q(a)$ for some property $Q : α → \Type$,
  then it is possible to \emph{transport} the path $p$ over $x$ to obtain a
  proof $\transport{Q}{p}{x} : Q(b)$. If $a$ and $b$ are types and $x : a$, we
  write $\coerce{p}{x} ≡ \transport{λx.x}{p}{x} : b$.

  Functions $f : α → β$ in HoTT are \emph{functorial}, meaning that paths $p : a
  = b$ on $α$ can be promoted to paths $\ap[_f]{p} : f a = f b$. An
  \emph{equivalence} $f : α ≅ β$ between types $α$ and $β$ consists of a pair of
  functions $f : α → β$ and $f^{-1} : β → α$, along with proofs $η : \prod_a
  f^{-1}(f a) = a$ and $ε : \prod_b f (f^{-1} b) = b$ such that $\prod_a
  \ap[_f]{η_a} = ε_{f~a}$~\cite[Definition 4.2.1]{hott2013}. The
  \emph{univalence axiom} states that an equivalence $f : α ≅ β$ between types
  can be treated as a path $\univ{f} : α = β$, such that $\coerce{(\univ{f})}{a}
  = f a$:
  \[
    \UNIV : \prod_{α,β} (α ≅ β) ≅ (α = β).
  \]

\subsection{Main idea---unitaries as paths}
\label{sec:LTypeDefn}

The core idea of this work is to encode the unitary operators used in quantum
computing in the higher inductive structure of quantum types.

We start by defining unitary matrices in homotopy type theory. Let $ℂ$ be a type of complex numbers in HoTT.\footnote{The precise formulation of the complex numbers in HoTT is not relevant for this work; for concreteness we can pick a representation based on the
Dedekind reals~\citep[Chapter 11]{hott2013}. But we never need to define functions out of $ℂ$, and so all that we require is that basic computational properties, such as arithmetic and the complex conjugate, are valid.} For any finite types
$α,β : \FinType$, let $\Matrix(α,β)$ be the set of complex-valued
$2^{|α|} \times 2^{|β|}$ matrices.\footnote{Here we use $|α|$ to refer to the size of the finite type $α$. The fact that $\Matrix(α,β)$ is a set
  means that for any two paths $p₁,p₂ : A = B$ between matrices
  $A,B : \Matrix(α,β)$, it is the case that $p₁ = p₂$.} We write $I$ for the
identity matrix, $A B$ for matrix multiplication, $A^\dagger$ for the conjugate
transpose of $A$, $A ⊗ B$ for the tensor product, and $A ⊕ B$ for the block
matrix $\left( \begin{smallmatrix}A & 0 \\ 0 & B \end{smallmatrix} \right)$.

A unitary matrix $\UMatrix(α,β)$ is a matrix $U : \Matrix(α,β)$ such that its
conjugate transpose is its own inverse: $U^\dagger U = U U^\dagger = I$. Unitary
transformations $\UMatrix$ form a groupoid---a category whose objects are finite
types, and whose morphisms $\UMatrix(α,β)$ are all invertible. 

In this paper we encode unitary matrices in a quantum language by taking the
types of our quantum language to be elements of~$\LType ≡
\quotient{\FinType}{\UMatrix}$, the \emph{groupoid quotient} of
$\UMatrix$.\footnote{ \cref{sec:qtype} extends the set quotient type on
  equivalence relations $α/R$ to groupoid quotients $\quotient{α}{G}$.} The
intuition is that qubits are represented as the type $[\Bool]_\UMatrix$ of
two-dimensional vector spaces, and unitary transformations $U : \UMatrix(α,β)$
are encoded as paths of type $[α]_\UMatrix = [β]_\UMatrix$. For quantum types
$σ$ and $τ$, we write $𝒰(σ,τ)$ for the unitary path type $σ = τ$.

Encoding unitaries as paths has two important consequences. First, there is no
need for explicit syntax for applying a unitary transformation; unitary
application $U ♯ e$ is defined to be $\transport{}{U}{e}$, the result of
transporting the path $U : σ = τ$ over a term $Γ ⊢ e : σ$. Second, many of the
structural axioms on unitaries can now be proven by path induction. For example,
consider the following statement:
\begin{proposition}
  Suppose $Γ ⊢ e : σ$. Then, for $U : σ = τ$ and $V : τ = ρ$ we have
  $ 
    V ♯ (U ♯ e) = (V ∘ U) ♯ e.
  $ 
\end{proposition}
\begin{proof}
  By path induction over $V$. If $V$ is the trivial path by reflexivity on $σ$,
  written $1_σ$, then $V ∘ U = U$. By the definition of 
  $\TRANSPORT$, for all $x$ we have $1 ♯ x = x$. So
  $
    1 ♯ (U ♯ e) = U ♯ e = (1 ∘ U) ♯ e.
  $
\end{proof}

Crucially, it is \emph{not} possible to prove the following false statement:
\begin{proposition}[False]
  Let $Γ ⊢ e : σ$ and $U : σ = σ$. Then $U ♯ e = e$.
\end{proposition}
Path induction only applies on proofs $a = b$ when at least one of $a$ or $b$ is
a free variable, so it does not apply here. In fact, the statement is
false---\cref{prop:X-intro-elim} will show that $X ♯ \ket{0} = \ket{1}$, but it
is not the case that $\ket{0} = \ket{1}$ due to the soundness of the
denotational semantics (\cref{thm:soundness}).



\section{A quantum term calculus}
\label{sec:lambda}
In this section we present a specification of an embedded linear quantum
language. The embedding, closely related to the linearity monad embedding~\cite{paykin2017a}, allows us
to use non-linear data in a natural way in the style of linear/non-linear type
theory~\citep{benton1995}. In \cref{sec:types} we will implement this
language in HoTT by encoding unitary transformations as paths.


Quantum types $σ,τ : \LType$ consist of linear pairs $σ ⊗ τ$ and sums $σ ⊕ τ$,
as well as host language types $\Lower{α}$. Host types $α : \Type$ are also
called \emph{classical}, in contrast to quantum types. We choose to restrict our
semantics to finite-dimensional vector spaces, so we insist that $α$ be
finite for it to be used in a quantum type. That is, for any
finite host language type $α : \FinType$, there is a quantum type $\Lower{α}$
containing (superpositions of) host language values of type $α$.
\[ \begin{array}{c}
    \inferrule*
    {α : \FinType}
    {\Lower{α} : \LType}
  \qquad
    \inferrule*
    {σ : \LType \\ τ : \LType}
    {σ ⊗ τ : \LType}
  \qquad
    \inferrule*
    {σ : \LType \\ τ : \LType}
    {σ ⊕ τ : \LType}
\end{array} \]

A typing context $Γ : \Ctx$ is a finite map from linear variables $\Var$ to
$\LType$s. For $Γ₁$ and $Γ₂ : \Ctx$, we write $Γ₁,Γ₂$ for the disjoint merge of
$Γ₁$ and $Γ₂$, which is only defined when $Γ₁$ and $Γ₂$ are disjoint. We write
$∅$ for the empty typing context and $x:σ$ for the singleton typing context.

Linear (quantum) expressions are given by the type $\LExp{Γ}{σ}$, where $Γ$ is a
typing context and $σ$ is a quantum type. $\QEXP$ represents a typing
judgment; we sometimes write $Γ ⊢ e : σ$ for $e : \LExp{Γ}{σ}$.

The type $\LExp{Γ}{σ}$ is defined inductively by the rules in \cref{fig:LNL}.
Most of the constructions are standard for a linear lambda calculus.
The exception is the constructors of the non-linear data type $\Lower{α}$, which we explain in more detail here.

The introduction rule for $\Lower{α}$ says that for any $a:α$ in the host
type theory, there is a linear expression $\put{a}$ of quantum type $\Lower{α}$
that does not use any linear variables. Terms of type $\Lower{α}$
correspond to quantum states: vectors in an $|α|$-dimensional,
complex-valued vector space, where the $i$th element $a_i : α$ corresponds to
a vector $(0,…,1,…,0)^\dagger$ with  $1$ at index $i$ and $0$ elsewhere.\footnotemark

\footnotetext{This intuition will be formalized in \cref{sec:semantics}.}

The elimination rule for $\Lower{α}$ has the form $e ⸖ f$, where $e$ is a
quantum term of type $\Lower{α}$ and $f$ is a host-level function from values of
type $α$ to quantum expressions. The infix constructor $⸖$ is pronounced
``let-bang'', in reference to the linear logic operation $!$, pronounced
``bang''.
Intuitively, the expression $e ⸖ f$ \emph{measures}\footnotemark~ the quantum expression $e$,
resulting in a value of type $α$, and then applies $f$ to that value. We will
sometimes write $\letin{!x}{e}{e'}$ for $e ⸖ λx.e'$ when it is clear that $x$ is
a host-level variable, as opposed to a linear variable bound by the linear
typing context.

\footnotetext{The basis in which this measurement occurs corresponds to the order of the finite type $α$.}

\begin{figure*}
\[ \begin{array}{c}
    \inferrule*[right=var]
    {Γ = x:σ}
    {x : \LExp{Γ}{σ}}
  \qquad\qquad
    \inferrule*[right=let]
    { e : \LExp{Γ}{σ} \\
      e' : \LExp*{Γ',x:σ}{τ}
    }
    { \letin{x}{e}{e'} : \LExp*{Γ,Γ'}{τ} }
  \\ \\
    \inferrule*[right=⊗-I]
    { e₁ : \LExp{Γ₁}{σ₁} \\
      e₂ : \LExp{Γ₂}{σ₂} }
    { (e₁,e₂) : \LExp*{Γ₁,Γ₂}*{σ₁ ⊗ σ₂} }
  \qquad
    \inferrule*[right=⊗-E]
    { e : \LExp{Γ}*{σ₁ ⊗ σ₂} \\
      e' : \LExp*{Γ',x₁:σ₁,x₂:σ₂}{τ} }
    { \letin{(x₁,x₂)}{e}{e'} : \LExp*{Γ,Γ'}{τ} }
  \\ \\
    \inferrule*[right=⊕-I]
    { e_i : \LExp{Γ}{σ_i} }
    { ι_i e_i : \LExp{Γ}*{σ₁ ⊕ σ₂} } 
  \quad
    \inferrule*[right=⊕-E]
    { e : \LExp{Γ}*{σ₁ ⊕ σ₂} \quad
      e₁ : \LExp*{Γ',x₁ : σ₁}{τ} \quad
      e₂ : \LExp*{Γ',x₂ : σ₂}{τ} }
    {\caseof{e}{ι₁ x₁ → e₁}{ι₂ x₂ → e₂} : \LExp*{Γ,Γ'}{τ} }
  \\ \\ 
    \inferrule*[right=lower-I]
    { a : α}
    { \put{a} : \LExp{∅}*{\Lower{α}} }
  \qquad
    \inferrule*[right=lower-E]
    { e : \LExp{Γ}*{\Lower{α}} \\
      f : α → \LExp{Γ'}{τ} }
    { e ⸖ f : \LExp*{Γ,Γ'}{τ} }
\end{array} \] 
\caption{An embedded linear/non-linear type system.}
\label{fig:LNL}
\end{figure*}

\subsection{Equational theory}

The behavior of this calculus is given as an equational theory, so that we can justify our axioms of quantum computing.
We write $e\{e₁/x₁,…,eₙ/xₙ\}$ for the simultaneous capture-avoiding substitution
of the $eᵢ$'s for the linear variables $xᵢ$ in $e$, and we write $e₁ ∼_α e₂$ for
the usual notion of $α$ equivalence. \cref{fig:LNL-beta} shows the $β$ equivalences for the language, which we write as $e₁ ∼_β e₂$.

\begin{figure}
\begin{align*}
    \letin{x}{e}{e'} &∼_β e'\{e/x\}                         \tagsc{β-let} \\
    \letin{(x₁,x₂)}{(e₁,e₂)}{e'} &∼_β e'\{e₁/x₁,e₂/x₂\}     \tagsc{β-⊗} \\
    \caseof{ιᵢ e}{ι₁ x₁ → e₁}{ι₂ x₂ → e₂} &∼_β eᵢ\{e / xᵢ\} \tagsc{β-⊕} \\
    \put{a} ⸖ f &∼_β f a                                    \tagsc{β-lower}
\end{align*}
\caption{Linear/non-linear $β$  equivalences}
\label{fig:LNL-beta}
\end{figure}

Eta equivalence, written $e₁ ∼_η e₂$, is allowed for product types but not, in
general, for sums $σ ⊕ τ$ or classical types $\Lower{α}$. Semantically, case
analysis for sums and $⸖$ for $\Lower{α}$ perform quantum measurement (see
\cref{sec:measurement}), so a term $e$ of type $σ ⊕ τ$ is \emph{not}
semantically equivalent to its $η$-expanded version,
$\caseof{e}{ι₁ x → ι₁ x}{ι₂ x → ι₂ x}$.

However, $η$ expansion for the multiplicative product and unit type $\Lower{()}$
are admissible, since they do not encode classical information. In fact, for the
unit type we have an even stronger property---any two values of unit type are
equivalent. This reflects the fact that the unit type is terminal in the
category of density matrices in which we define a denotational semantics in
\cref{sec:semantics}.
\[\begin{array}{c}
    \inferrule*[right=η-⊗]
    {Γ ⊢ e : σ₁ ⊗ σ₂}
    {e ∼_η \letin{(x₁,x₂)}{e}{(x₁,x₂)}}
  \qquad
    \inferrule*[right=η-()]
    {Γ ⊢ e₁ : \Lower{()} \\ Γ ⊢ e₂ : \Lower{()}}
    {e₁ ∼_η e₂}
\end{array} \]
Recall that we write $Γ ⊢ e : τ$ for $e : \LExp{Γ}{τ}$. Notice that the usual $η$ rule for $\Lower{()}$---that $e ∼_η \letin{()}{e}{()}$---can be derived from \textsc{η-()}.


Commuting conversions, written $e ∼_{cc} e'$, describe how elimination forms can
move within an expression~\citep[Chapter 10]{girard1989}. Rules of this form,
shown in \cref{fig:cc}, are common for linear lambda calculi~\citep{staton2013}.

\begin{figure*}
\[ \begin{array}{c}
    \inferrule*[right=cc-let]
    { Γ ⊢ e : σ \\
      Γ',y:σ ⊢ e' : τ \\
      Γ₀,x : τ ⊢ e₀ : τ₀ 
    }
    {e₀\{\letin{y}{e}{e'}/x\} ∼_{cc} \letin{y}{e}{e₀\{e'/y\}}}
  \\ \\
    \inferrule*[right=cc-⊗]
    {Γ ⊢ e : σ₁ ⊗ σ₂ \\
      Γ',y₁:σ₁,y₂:σ₂ ⊢ e' : τ \\
      Γ₀,x : τ ⊢ e₀ : τ₀ 
    }
    {e₀\{\letin{(y₁,y₂)}{e}{e'}/x\} ∼_{cc} \letin{(y₁,y₂)}{e}{e₀\{e'/x\}}}
  \\ \\
    \inferrule*[right=cc-⊕]
    { Γ ⊢ e : σ₁ ⊕ σ₂ \\
      Γ',y₁:σ₁ ⊢ e₁ : τ \\
      Γ',y₂:σ₂ ⊢ e₂ : τ \\
      Γ₀,x:τ ⊢ e₀ : τ₀ }
    { e₀\{\caseof{e}{ι₁ y₁ \shortrightarrow e₁}
                    {ι₂ y₂ \shortrightarrow e₂} / x\} 
      ∼_{cc} \caseof{e}{ι₁ y₁ \shortrightarrow e₀\{e₁/x\}}
                       {ι₂ y₂ \shortrightarrow e₀\{e₂/x\}} }
  \\ \\
    \inferrule*[right=cc-Lower]
    { Γ ⊢ e : \Lower{α} \\
      \prod_{a:α} Γ' ⊢ f a : τ \\
      Γ₀,x:τ ⊢ e₀ : τ₀
    }
    { e₀\{e ⸖ f / x\} ∼_{cc} e ⸖ λ a. e₀ \{f a / x \} }
\end{array} \]
\caption{Commuting conversion rules}
\label{fig:cc}
\end{figure*}

\subsection{Linear functions} 
The quantum language described so far does not include higher-order functions,
as the physical interpretation of higher-order functions in quantum mechanics is
not entirely clear~\citep{valiron2008,malherbe2010,hasuo2011,pagani2014}. However, we can encode first-order functions in our host language in a
data type \coq{σ ⊸ τ} (pronounced $σ$ ``lolli'' $τ$) that represents quantum
computations with input $σ$ and output $τ$. The type $σ ⊸ τ$ is defined
inductively by a single constructor, $\SUSPEND$, that takes an expression of
type $τ$ with a single argument of type $σ$:
\[ \SUSPEND : \prod_x \LExp*{x:σ}{τ} → (σ ⊸ τ). \] 
The eliminator says that for any $k : \prod_x \LExp*{x:σ}{τ} → β$ there is a
function $\REC_⊸^k : (σ ⊸ τ) → β$ such that %
$\rec*[_⊸^k]{x~e} ≡ k~x~(fx)$.\footnotemark~ To apply a linear function 
$g : σ ⊸ τ$ to an argument $Γ ⊢ e : σ$, define $Γ ⊢ \FORCE~g~e : τ$ using
$\REC_⊸^k$ with $k ≡ λ x~e₀.e₀\{e/x\}$, so that 
$\force*{\suspend{x~e'}}~e ≡ e'\{e/x\}$.

\footnotetext{The induction principle says that for any $P : (σ ⊸ τ) → \Type$,
  given a proof $p : \prod_x \prod_{e : \LExp*{x:σ}{τ}} P~(\suspend{x~e})$,
  there is a function $\IND_⊸^p : \prod_{f : σ ⊸ τ} P~f$ such that
  $\IND_⊸^p(\suspend{x~e}) ≡ p~x~e$.}

The quantum identity function is $\suspend*{x.x}$, and the function that swaps
the elements of a pair is $\suspend*{(x,y).(y,x)}$ of type $σ ⊗ τ ⊸ τ ⊗ σ$.


\subsection{Quantum data, and measurement as case analysis}
\label{sec:measurement}

A qubit is a quantum type with two basis elements, $\ket{0}$ and $\ket{1}$, so we encode qubits as $\Qubit ≡ \Lower{\Bool}$. Initialization and measurement of qubits is straightforward:
\[ \begin{aligned}
    \INIT &: \Bool → \LExp{∅}{\Qubit} \\
        &≡ λ b.~ \put{b} 
    \end{aligned} \qquad\qquad \begin{aligned}
    \MEAS &: \Qubit ⊸ \Lower{\Bool} \\
        &≡ \suspend*{λ x.~\letin{!b}{x}{\put{b}}}
\end{aligned} \]
On first glance, these definitions appear not to be doing
anything---$\force*{\meas{e}}$ in particular is just the $η$ expansion of $e$.
But this highlights a critical semantic fact of our system: \emph{case analysis
  performs quantum measurement}. This has a number of consequences for the
theory of the language, including the fact that $η$ expansion is not sound in
general: a measured qubit is \emph{not} equivalent to an unmeasured one.

By choosing to encode measurement as case analysis, we open the door to a very
expressive quantum theory. For example, the type $\Lower*{\Bool \times \Bool}$ is
equivalent to the two-qubit system
$\Lower{\Bool} ⊗ \Lower{\Bool}$. Qutrits (a base-3 quantum system) can be encoded as $\Lower*{() + () + ()}$, and finite, length-indexed lists of qubits can be encoded as $\Lower*{\Vec{n}{\Bool}}$.

\subsection{Unitary transformations.}
\label{sec:statonTheory}

The remaining components of the language are unitary
transformations $𝒰(σ,τ)$.
A unitary $U : 𝒰(σ,τ)$ can be applied to an expression of type $σ$ to produce
an expression of type $τ$:
\[
    \inferrule*
    { U : 𝒰(σ,τ) \\
      e : \LExp{Γ}{σ} }
    { \unitary{U}{e} : \LExp{Γ}{τ} }
\]
\cref{sec:types} will show how to derive the unitaries in the lambda calculus
described so far; this section will spell out a specification of the properties
we expect to hold of the quantum fragment.

Following Staton~\cite{staton2015}, we focus on four main ways to combine unitaries:
if $U : 𝒰(q,r)$ and $V : 𝒰(q',r')$, then $U ⊗ V : 𝒰(q ⊗ q',r ⊗ r')$ is the
tensor product of $U$ by $V$, and $U ⊕ V : 𝒰(q ⊕ q',r ⊕ r')$ is the block
matrix. In addition, unitaries form a groupoid: there is an identify unitary,
written $1$; unitaries are subject to composition, written $V ∘ U$; and they are
invertible, written $U^\dagger$. Staton proves that all unitary matrices can be
constructed from 1-qubit unitaries with the direct sum and tensor
product~\citep{staton2015}.



The equational theory of unitaries is divided into three classes. First, the
``structural'' axioms, shown in \cref{fig:structural-axioms}, characterize the
how unitaries interact with syntactic forms of the language. For example,
\cref{eqn:U-tensor-intro} describes how the tensor product $U₁ ⊗ U₂$ distributes
over pairs.

\begin{figure*}
\begin{minipage}{\textwidth}
\begin{align}
    \unitary{(U₁ ⊗ U₂)}{(e₁,e₂)} 
        &≈ (\unitary{U₁}{e₁}, \unitary{U₂}{e₂}) 
        \label{eqn:U-tensor-intro} \tagsc{U-⊗-intro} \\
    \letin{(x₁,x₂)}{(U₁ ⊗ U₂) ♯ e}{e'} \notag \\
        ≈ \letin{(y₁,y₂)}{e}{&e'\{ U₁ ♯ y₁ / x₁, U₂ ♯ y₂ / x₂\}}
        \tagsc{U-⊗-elim} \\
    \unitary{U}{(\letin{(x₁,x₂)}{e}{e'})} 
        &≈ \letin{(x₁,x₂)}{e}{\unitary{U}{e'}} 
        \tagsc{U-⊗-comm} \\
    \notag \\
    \unitary{(U₁ ⊕ U₂)}{(ι₁ e)} 
        &≈ \unitary{U₁}{e} 
        \tagsc{U-⊕-intro₁} \label{eqn:U-oplus-intro0} \\
    \unitary{(U₁ ⊕ U₂)}{(ι₂ e)} 
        &≈ \unitary{U₂}{e} 
        \tagsc{U-⊕-intro₂} \label{eqn:U-oplus-intro2} \\
    \caseof{(U₁ ⊕ U₂) ♯ e}{ι₁ x₁ → e₁}{ι₂ x₂ → e₂}  \notag \\
        ≈ \caseof{e}{ι₁ y₁ → e₁\{U₁ ♯ y₁ / x₁\}}{&ι₂ y₂ → e₂\{U₂ ♯ y₂ / x₂\}} 
        \tagsc{U-⊕-elim} \label{eqn:U-oplus-elim} \\
    \unitary{U}{(\caseof{e}{ι₁ x₁ → e₁}{ι₂ x₂ → e₂})} \notag \\
        ≈ \caseof{e}{ι₁ x₁ → \unitary{U}{e₁}}
                     {&ι₂ x₂ → \unitary{U}{e₂}} 
        \tagsc{U-⊕-comm} \\
    \notag \\ 
    \unitary{U}{(e ⸖ f)} 
        &≈ e ⸖ λ x → \unitary{U}{(f x)} 
        \tagsc{U-Lower-comm} \\
    U ♯ e ⸖ λ \_.e'
        &≈ e ⸖ λ \_.e'
        \tagsc{U-Lower-elim} \label{eqn:U-Lower-elim} 
\end{align}
\end{minipage}
\caption{Structural axioms. The relation $e ≈ e'$ is the union of $α$, $β$, $η$,
  and commuting conversion relations, along with the quantum-specific axiom that
  will be defined in \cref{sec:axioms}.}
\label{fig:structural-axioms}
\end{figure*}

Second, the axioms in \cref{fig:groupoid-axioms} characterize that unitaries
form a groupoid.

\begin{figure}
\begin{align}
    \unitary{U}{(\unitary{V}{e})} &≈ \unitary{(U ∘ V)}{e} 
        \tagsc{U-compose} \label{ax:U-compose} \\
    \unitary{I}{e} &≈ e 
        \tagsc{U-I} \label{ax:U-I} \\
    U^\dagger ♯ U ♯ e &≈ e
        \tagsc{U-$\dagger$} \label{ax:U-dag}
\end{align}
\caption{Groupoid axioms}
\label{fig:groupoid-axioms}
\end{figure}

The third set of axioms describe how certain unitaries interact with
initialization and measurement, for instance those shown in
\cref{fig:semantic-axioms}. Such unitaries are completely defined by
isomorphisms on their basis sets, which we call \emph{unitary equivalences}
$σ ⇆ τ$, and define formally in \cref{sec:axioms}.



Every unitary equivalence $f$ can be lifted to a unitary transformation
$\widetilde{f} : 𝒰(σ,τ)$. For example, for the equivalence $\swap :
\prod_{σ₁,σ₂} σ₁ ⊗ σ₂ ⇆ σ₂ ⊗ σ₁$, it should be the case that $\widetilde{\swap}
♯ (e₁,e₂) ≈ (e₂,e₁)$ as shown in \cref{fig:semantic-axioms}. We call $(e₁,e₂)$
the \emph{partial initialization} of the quantum system $Π_{σ_1,σ_2} σ₁ ⊗ σ₂$,
reflected by the fact that $\swap$ quantifies over all types $σ₁$ and $σ₂$.
Partial initialization and its counterpart, partial measurement, precisely
characterize the behavior of unitary equivalences $f : σ ⇆ τ$:
\begin{align}
    \widetilde{f} ♯ \init[_σ]{b} &≈ \init*[_τ]{f b} 
        \tagsc{U-intro} \\
    \matchwith{τ}{(\widetilde{f} ♯ e)}{g} &≈ \matchwith{σ}{e}{g ∘ f}.
       \tagsc{U-elim}
\end{align}

\begin{figure*}
\begin{align}
    X ♯ \init{b} &≈ \init*{¬ b} 
        \tagsc{X-intro} \label{ax:X-intro} \\
    \letin{!x}{\meas*{X ♯ e}}{e'} &≈ \letin{!y}{\meas{e}}{e'\{¬ y / x\}}
        \tagsc{X-elim} \label{ax:X-elim} \\
    \notag \\
    \SWAP ♯ (e₁,e₂) &≈ (e₂,e₁) 
        \tagsc{\SWAP-intro} \label{ax:SWAP-intro} \\
    \letin{(x,y)}{\SWAP ♯ e}{e'} &≈ \letin{(y,x)}{e}{e'} 
        \tagsc{\SWAP-elim} \label{ax:SWAP-elim} \\
    \notag \\
    \DISTR ♯ (\init{b},e) &≈ \IfThenElse{b}{ι₂~e}{ι₁~e}
        \tagsc{\DISTR-intro} \label{ax:DISTR-intro} \\
    \caseof*{\DISTR ♯ e}{ι₁ z₁ → e₁}{ι₂ z₂ → e₂} 
        &≈ \letin{(!b,y)}{e}{(\init{b},e)}
        \tagsc{\DISTR-elim} \label{ax:DISTR-elim}
\end{align}
\caption{Examples of axioms describing the behavior of the unitaries
  $X : 𝒰(\Qubit,\Qubit)$, $\SWAP : 𝒰(σ₁ ⊗ σ₂, σ₂ ⊗ σ₁)$, and
  $\DISTR : 𝒰(\Qubit ⊗ τ, τ ⊕ τ)$.}
\label{fig:semantic-axioms}
\end{figure*}




\section{Deriving equational rules in HoTT}
\label{sec:types}
The goal of this section is to encode unitaries in 
quantum types, to minimize the number of axioms needed to recover the
equational theory described in the previous section. As described in
\cref{sec:LTypeDefn}, we do this by encoding unitaries in the
groupoid quotient $\LType ≡ \FinType /₁ \UMatrix$.

\cref{sec:quotient1,sec:qtype} have been formalized in Coq using the HoTT library.\footnote{The formalization is available on github: \url{https://github.com/jpaykin/GroupoidQuotient}. It uses the HoTT library, an extention to Coq that supports homotopy type theory and higher inductive types~\citep{bauer2017}.}

\subsection{Groupoid quotient as a higher inductive type}
\label{sec:quotient1}

\begin{definition}[Sojakova~\cite{sojakova2015}, Section~4.3] \label{defn:groupoid-quotient}
  If $G$ is a groupoid with objects $α$, then the \emph{groupoid quotient} of
  $G$, written $\quotient{α}{G}$, is a higher inductive type with the
  following constructors:
  \[ \begin{aligned}
    \POINT &: α → \quotient{α}{G} \\ \\
    \CELL &: \prod_{a,b} G(a,b) → \point{a} = \point{b}
  \end{aligned} \qquad \begin{aligned}
    \CELLCOMPOSE &: \prod_{f,g} \cell*{g ∘ f} = \cell{g} ∘ \cell{f} \\
    \ISONETYPE &: \onetype*{\quotient{α}{G}}
  \end{aligned}\]
  The fact that $\quotient{α}{G}$ is a 1-type means that, for
  $x,y : \quotient{α}{G}$, if $f,g : x = y$ and $p,q : f=g$, then $p = q$.

  The induction principle states: for a predicate $P$ on
  $\quotient{α}{G}$, there is a proof $\kwfont{ind}_P$ of $\prod x, P x$,
  provided
  \begin{enumerate}[-]
    \item For all $x$,  $P x$ is a 1-type;
    \item For all $a : α$, there is a proof $\PPoint{a}$ of 
      $P (\point{a})$;
    \item For all $f : G(a,b)$, there is a proof $\PCell{f}$
        that $\transport{P}{(\cell{f})}{(\PPoint{a})} = \PPoint{b}$; 
    \item For $f : G(a,b)$ and $g : G(b,c)$, the diagram in \cref{fig:quotient-ind} commutes.
  \end{enumerate}

    \noindent
    Furthermore, $\kwfont{ind}_P$ satisfies the following computation laws:
    \begin{align*}
        \kwfont{ind}_P(\point{a}) ≡ \PPoint{a} 
      \qquad\text{and}\qquad
        \apd{\kwfont{ind}_P}(\cell{f}) ≡ \PCell{f}
    \end{align*}
    where, for $f : \prod (x:α), P(x)$ and $p : a = b$ at type $α$, we have
    $\apd{f}(p) : \transport{P}{p}{(f a)} = f b$.
\end{definition}

\begin{figure*}
    \begin{center}
    \begin{tikzpicture}[scale=0.6,every node/.style={scale=0.85}] 
        \node (a) 
            {$\transport{P}{(\cell*{g ∘ f})}{(\PPoint{a})}$};
        \node[below=of a] (b) 
            {$\transport{P}{(\cell{g} ∘ \cell{f})}{(\PPoint{a})}$};
        \node[below right= and -2.5cm of b] (c)
            {$\transport{P}{(\cell{g})}{\left(
          \transport{P}{(\cell{f})}{(\PPoint{a})}\right)}$};
        \node[above right= and -2.5cm of c] (d)
            {$\transport{P}{(\cell{g})}{(\PPoint{b})}$};
        \node[above=of d] (e)
            {$\PPoint{c}$};

        \draw[double distance=2pt] (a) 
             -- node[anchor=east] {$\ap{}{(\cellcompose{f}{g})}$} 
             (b);

        \draw[double distance=2pt] (a)
             -- node[anchor=south] {$\PCell{g ∘ f}$}
             (e);

        \draw[double distance=2pt] (b)
             -- 
             (c);
        
        \draw[double distance=2pt] (c)
             -- node[anchor=north west] {$\ap{}{\PCell{f}}$}
             (d);

        \draw[double distance=2pt] (d)
             -- node[anchor=west] {$\PCell{g}$}
             (e);
    \end{tikzpicture}
    \end{center}
\caption{A condition for quotient induction.}
\label{fig:quotient-ind}
\end{figure*}

\subsection{\texttt{QType} as a groupoid quotient}
\label{sec:qtype}

Define $\LType$ to be the groupoid quotient of $\UMatrix$:
$\LType ≡ \quotient{\FinType}{\UMatrix}$. Intuitively, for $σ,τ :\LType$, the type $σ = τ$ corresponds to unitary transformations from $σ$ to $τ$.
The groupoid quotient ensures that the identity and inverse of paths
correspond to the appropriate operations on matrices.

\begin{proposition} \label{prop:cellI} Let $I : \UMatrix(α,α)$ be the
  identity matrix on $α$. Then $\cell{I} = 1_{\point{α}}$.
\end{proposition}
\begin{proof}
  Since $I = I ∘ I$, by the compositionality of $\CELL$ we know that
  $\cell{I} = \cell{I} ∘ \cell{I}$. But for any path $p : x = x$, if $p ∘ p = p$
  then $p$ must be $1_x$.
\end{proof}

\begin{proposition} \label{prop:celldag}
  Let $U : \UMatrix(α,β)$. Then $(\cell{U})^{-1} = \cell{U^\dagger}$.
\end{proposition}
\begin{proof}
  By the compositionality of $\CELL$ and \cref{prop:cellI}, 
  $ \cell{U} ∘ \cell{U^\dagger} = \cell*{U ∘ U^\dagger} = \cell{I} = 1. $
\end{proof}


Next we need to define the type formers $\LOWER$, $⊗$, and $⊕$. We take $\Lower{α}$ to be $(\point{α}) : \LType$, and $⊗$ and $⊕$ are defined by
quotient induction.
To define $⊗ : \LType → \LType → \LType$, we apply a variant of the quotient
recursion principle on two variables.
\begin{lemma}
  Let $G_1$ and $G_2$ be groupoids with objects $α_1$ and $α_2$ respectively.
  Let $β$ be a 1-type, and let $f : α_1 → α_2 → β$ be a function, with
  \[
    f^{\CELL} : \prod_{x_1,y_1 : α_1} \prod_{x_2,y_2 : α_2} G_1(x_1,y_1) → G_2(x_2,y_2) → f~x_1~x_2 = f~y_1~y_2
  \]
  such that $ f^{\CELL}~(h_1 ∘ g_1)~(h_2 ∘ g_2) = f^{\CELL}~h_1~h_2 ∘ f^{\CELL}~g_1~g_2$ for all $g_i : G_i(x_i,y_i)$ and $h_i:G_i(y_i,z_i)$, 

  Then there is a function $f^{\REC2} : \quotient{α_1}{G_1} → \quotient{α_2}{G_2} → β$ with
  \begin{align*}
    f^{\REC2}~(\point{x_1})~(\point{x_2}) &= f~x_1~x_2
  \qquad\text{and}\qquad
    \apTwo[_{f^{\REC2}}]{\cell{g_1}}{\cell{g_2}} = f^{\CELL}~g_1~g_2
  \end{align*}
  for all $x_1,y_1:α_1$, $x_2,y_2:α_2$, $g_1:G_1(x_1,y_1)$, and $g_2:G_2(x_2,y_2)$, where 
    \[ \apTwo{~} : \prod_{f : A → B → C} \prod_{a_1,a_2:A} \prod_{b_1,b_2:B} a_1 = a_2 → b_1 = b_2 → f~a_1~b_1=f~a_2~b_2. \]
\end{lemma}

Thus, to define $⊗$, it suffices to define how
it acts on points and cells, and then show that it is bilinear. We
define
$
    \point{α₁} ⊗ \point{α₂} ≡ \point*{α₁ \times α₂}.
$
If $U : \UMatrix(α,α')$ and $V : \UMatrix(β,β')$ then we have
$\cell*{U ⊗ V} : \point*{α \times β} = \point*{α' \times β'}$. The remaining
condition is  that 
\[
    \cell*{U₂ ⊗ V₂} ∘ \cell*{U₁ ⊗ V₁} = \cell*{(U₂ ∘ U₁) ⊗ (V₂ ∘ V₁)},
\]
which follows from the fact of linear algebra that
$
    (U₂ ∘ U₁) ⊗ (V₂ ∘ V₁) = (U₂ ⊗ V₂) ∘ (U₁ ⊗ V₁).
$

For $U : σ = τ$ and $U' : σ' = τ'$, we lift the tensor product to
$U ⊗ U' ≡ \ap[_⊗]{(U,U')}: σ ⊗ σ' = τ ⊗ τ'$. The computation principle
states that 
$
    \cell{U} ⊗ \cell{U'} = \cell*{U ⊗ U'}.
$

  A similar argument is used to define $⊕$. 

\subsection{Deriving the groupoid axioms}

The fact that unitaries are paths means that \cref{fig:groupoid-axioms}'s
groupoid axioms can be derived for free.


\begin{proposition}[\ref{ax:U-compose}] \label{prop:U-compose}
  Let $V : σ = τ$ and $U : τ = ρ$. Then
  $
    U ♯ (V ♯ e) = (U ∘ V) ♯ e.
  $
\end{proposition}
\begin{proof}
  By path induction on $V$. Since $1 ♯ e ≡ e$ and $U ∘ 1 = U$, the
  equation reduces to $U ♯ e=U♯e$.
\end{proof}

\begin{proposition}[\ref{ax:U-I}] \label{prop:U-I}
  If $e : \LExp{Γ}{σ}$ then $\cell{I} ♯ e = e$.
\end{proposition}
\begin{proof}
  Follows from \cref{prop:cellI} which states that $\cell{I} = 1$.
\end{proof}

\begin{proposition}[\ref{ax:U-dag}] \label{prop:U-dag}
  If $U : σ = τ$ and $e : \LExp{Γ}{σ}$ then $U^\dagger ♯ U ♯ e = e$.
\end{proposition}
\begin{proof}
  Follows from \cref{prop:U-compose} and the fact that, as matrices,
  $U^\dagger ∘ U = I$.
\end{proof}

\subsection{Deriving the structural axioms}

With one exception, the structural axioms from \cref{fig:structural-axioms} are
trivial by path induction, and we omit their proofs here. The exception is the behavior of \ref{eqn:U-Lower-elim}: when a qubit is measured but the result is not relevant to the rest of the computation.

\begin{proposition}
    \label{prop:U-Lower-distr}
    If $e : \LExp{Γ}{\Qubit}$ and $U : \Qubit = \Qubit$, then 
    \[ \letin{!\_}{\meas*{U ♯ e}}{e'} ≈ \letin{!\_}{\meas{e}}{e'}. \]
\end{proposition}
\begin{proof}
  It is not possible to do induction on $U$ here, since its endpoints are both
  fixed. However, \cref{prop:U-Lower-distr} follows from the $η$ rule for the
  unit type: for any two terms $e_1,e_2 : \LExp{Γ}*{\Lower{()}}$, 
  $e₁ ∼_η e₂$:
\begin{align*} 
    \letin{!\_}{\meas*{U ♯ e}}{e'}
    &∼ \letin{!\_}{\letin{!\_}{\meas*{U ♯ e}}{\put{()}}}{e'} \\
    &∼_η \letin{!\_}{\letin{!\_}{\meas{e}}{\put{()}}}{e'} \\
    &∼_η \letin{!\_}{e}{e'}  \qedhere
\end{align*}
\end{proof}


\section{Equivalence of unitaries}
\label{sec:axioms}
This section establishes the soundness of the unitary equivalences shown in
\cref{fig:semantic-axioms}. First, the ``not'' unitary $X$:

\begin{proposition}[\labelcref{ax:X-intro}~and~\labelcref{ax:X-elim}] \label{prop:X-intro-elim}
\[ \begin{aligned}
    \cell{X} ♯ \put{b} &= \put*{¬ b}  \\
  \end{aligned} \qquad\text{and}\qquad \begin{aligned}
    (\cell{X} ♯ e) ⸖ f &= e ⸖ λ b. f(¬ b)
\end{aligned} \]
\end{proposition}

The proof of this proposition relies on the following two lemmas, both proved by path induction:

\begin{lemma}\label{lem:put-letbang-commute}
  For any $f : α = β$ and $a : α$:
  \[\begin{aligned}
    \ap[_\POINT]{f} ♯ \put{a} &= \put*{\coerce{f}{a}}
    \end{aligned} \qquad\text{and}\qquad\begin{aligned}
    (\ap[_\POINT]{f} ♯ e) ⸖ g &= e ⸖ λ x.~g\left(\coerce{f}{x}\right).
  \end{aligned} \]
\end{lemma}

\begin{lemma}\label{lem:cell-transport}
  If $U : \UMatrix(α₁,α₂)$ and $H : α₂ = α₃$, then 
  \[
    \cell*{\transport{}{H}{U}} 
  =  \ap[_\POINT]{H} ∘ \cell{U}.
  \]
\end{lemma}

\begin{proof}[Proof of \cref{prop:X-intro-elim}]
  Instantiating \cref{lem:put-letbang-commute} with $(\univ{¬})$, it suffices
  to check that $\cell{X} = \ap*[_\POINT]{\univ{¬}}$. Notice that, as matrices,
  $X = \transport{\UMatrix(\Bool,-)}{(\univ{¬})}{I}$. Then
  \begin{align*}
    &\cell*{\transport{\UMatrix(\Bool,-)}{(\univ{¬})}{I}} \\
    &= \ap*[_\POINT]{\univ{¬}} ∘ \cell{I} 
        &&(\cref{lem:cell-transport}) \\
    &= \ap*[_\POINT]{\univ{¬}} 
        &&({\cref{prop:cellI}})
        \qedhere
  \end{align*}
\end{proof}


This technique does not extend to polymorphic equivalences such as
$\swap : \prod α β, α \times β = β \times α$. \cref{lem:put-letbang-commute} tells us how
$\SWAP ≡ \ap[_\POINT]{\swap}$ behaves on classical states:
$\SWAP♯\put{(a,b)} = \put{(b,a)}$. But \cref{ax:SWAP-intro} is an even stronger
statement: that $\SWAP ♯ (e₁,e₂) ∼_q (e₂,e₁)$ for any $e₁$ and $e₂$. Similarly,
the elimination form of \cref{lem:put-letbang-commute} tells us that measuring
both components of $\SWAP ♯ e$, where $e$ is a pair of qubits, is the same as
measuring $e$ and then swapping its arguments. However, \cref{ax:SWAP-elim}
doesn't measure both qubits; it only eliminates the pair:
\[
    \letin{(x,y)}{\SWAP ♯ e}{e'} ∼_q \letin{(x,y)}{e}{e'}.
\]

We can think of $\SWAP$'s behavior as acting on a state whose structure is only
\emph{partially} known, corresponding to the  polymorphism of its
underlying function $\swap$. Our solution is to define a sort of
\emph{partial initialization} and \emph{partial measurement} that generalizes
this notion for $\swap$ and other polymorphic paths.

\subsection{Partial initialization and measurement}

\begin{figure*}
\[
\begin{aligned}
    [X]^m &≡ m X \\
    [\Lower{α}]^m &≡ α \\
    [σ₁ ⊗ σ₂]^m &≡ [σ₁]^m \times [σ₂]^m \\
    [σ₁ ⊕ σ₂]^m &≡ [σ₁]^m + [σ₂]^m 
\end{aligned}
\quad
\begin{aligned}
    γ_X^m(x:\Var) &≡ x:\point*{m X} \\
    γ_{\Lower{α}}^m(a : α) &≡ ∅ \\
    γ_{σ₁ ⊗ σ₂}^m(b₁,b₂) &≡ γ_{σ₁}^m(b₁), γ_{σ₂}^m(b₂) \\
    γ_{σ₁ ⊕ σ₂}^m(\inl~b₁) &≡ γ_{σ₁}^m(b₁) \\
    γ_{σ₁ ⊕ σ₂}^m(\inr~b₂) &≡ γ_{σ₂}^m(b₂)
\end{aligned}
\quad
\begin{aligned}
    \init[_X^m]{x} &≡ x \\
    \init[_{\Lower{α}}^m]{a} &≡ \put{a} \\
    \init[_{σ₁ ⊗ σ₂}^m]{(b₁,b₂)} &≡ 
        \left(\init[_{σ₁}^m]{b₁}, \init[_{σ₂}^m]{b₂}\right) \\
    \init*[_{σ₀ ⊕ σ₁}^m]{\inl~b₀} &≡ ι₀ (\init[_{σ₀}^m]{b₀}) \\
    \init*[_{σ₀ ⊕ σ₁}^m]{\inr~b₁} &≡ ι₁ (\init[_{σ₁}^m]{b₁})
\end{aligned}
\]
\begin{align*}
    \matchwith{X}{e}{bs} &≡ (bs~x) \{e/x\} \qquad \text{where $x$ is fresh} \\
    \matchwith{\Lower{α}}{e}{bs} &≡ e ⸖ bs \\
    \matchwith{σ₁ ⊗ σ₂}{e}{bs} &≡
        \letin{(x₁,x₂)}{e}{\matchwith{σ₁}{x₁}
                          {λ b₁.~ \matchwith{σ₂}{x₂}{λ b₂.~bs(b₁,b₂)}}} \\
    \matchwith{σ₀ ⊕ σ₁}{e}{bs} &≡
        \casesof{e}{ι₀ x₀ → \matchwith{σ₀}{x₀}{λ b₀.~bs(\inl~b₀)}}
                  {ι₁ x₁ → \matchwith{σ₁}{x₁}{λ b₁.~bs(\inr~b₁)}} 
\end{align*}
\caption{Operations on open quantum types}
\label{fig:opentypes}
\end{figure*}

Consider quantum types with the addition of type variables $X : \LTypeVar$:
\[ σ ::= X ∣ \Lower{α} ∣ σ₁ ⊗ σ₂ ∣ σ₁ ⊕ σ₂. \] 
We call these \emph{open quantum types}.
Given a map $m : \LTypeVar → \Type$, we can define a basis set corresponding to
$σ$, written $[σ]^m$, as shown in \cref{fig:opentypes}.

Let $m : \LTypeVar → \Type$ and let $\BVar$ be the constant map $λ \_.\Var$.
Then every $b : [σ]^\BVar$ gives rise to a typing context $γ_σ^m(b)$ as well as
a term using these variables: if $Γ = γ_σ^m(b)$ then
$Γ ⊢ \init[_σ^m]{b} : \point{[σ]^m}$ is called \emph{partial
  initialization}, as defined in \cref{fig:opentypes}.

Open quantum types also dictate how to eliminate terms of type
$\point{[σ]^m}$, called \emph{partial measurement}:
\[
    \inferrule*
    { Γ ⊢ e : \point{[σ]^m} \\
      bs : \prod_{b:[σ]^\BVar} γ_σ^m(b),Γ' ⊢ - : τ }
    { Γ,Γ' ⊢ \match{σ}{e}{bs} : τ }
\]

A unitary equivalence $σ ⇆ τ$ of open quantum types is a proof that
$[σ]^m ≅ [τ]^m$ for every $m$. For example, the equivalence $X ⊗ Y ⇆ Y ⊗ X$ is
given by $λ m.~λ(x,y).(y,x)$.
\[
    σ ⇆ τ ≡ \prod_m [σ]^m ≅ [τ]^m.
\]

\begin{lemma}
   \label{lem:context-cohesion}
  If $f : σ ⇆ τ$ then for every $b : [σ]^\BVar$ there is a path 
  $γ_τ^m(f b) = γ_σ^m(b)$.
\end{lemma}

\subsection{Proof of \cref{lem:context-cohesion}}
\label{sec:normal-sigmas}
The proof of \cref{lem:context-cohesion} depends on the observation that open
type equivalence $σ ⇆ τ$ is equivalent to the inductively defined relation
$σ ≋ τ$ presented in \cref{fig:inductive-equiv}. It is easy to check that every
proof $f : σ ≋ τ$ corresponds to an equivalence $\hat{f} : σ ⇆ τ$, and it is
also easy to check that \cref{lem:context-cohesion} follows for
inductively-generated equivalences $f : σ ≋ τ$.

\begin{figure}
\[ \scriptsize 
  \begin{array}{c}
    \inferrule*[Right=refl]
    {~}
    {σ ≋ σ}
  \qquad\qquad
    \inferrule*[Right=symm]
    {σ₁ ≋ σ₂}
    {σ₂ ≋ σ₁}
  \qquad\qquad
    \inferrule*[Right=trans]
    {σ₁ ≋ σ₂ \\ σ₂ ≋ σ₃}
    {σ₁ ≋ σ₃}
  \\ \\
    \inferrule*[Right=cong$_⊗$]
    {σ₁ ≋ σ₂ \\ τ₁ ≋ τ₂}
    {σ₁ ⊗ τ₁ ≋ σ₂ ⊗ τ₂}
  \qquad\qquad
    \inferrule*[Right=cong$_⊕$]
    {σ₁ ≋ σ₂ \\ τ₁ ≋ τ₂}
    {σ₁ ⊕ τ₁ ≋ σ₂ ⊕ τ₂}
\end{array} \]
\begin{align*} \scriptsize
    σ₁ ⊗ σ₂ &≋ σ₂ ⊗ σ₁ \tag{$\SWAP_⊗$} \\
    σ₁ ⊕ σ₂ &≋ σ₂ ⊕ σ₁ \tag{$\SWAP_⊕$}\\
    σ₁ ⊗ (σ₂ ⊗ σ₃) &≋ (σ₁ ⊗ σ₂) ⊗ σ₃ \tag{$\ASSOC_⊗$}\\
    σ₁ ⊕ (σ₂ ⊕ σ₃) &≋ (σ₁ ⊕ σ₂) ⊕ σ₃ \tag{$\ASSOC_⊕$}\\
    σ₁ ⊗ (σ₂ ⊕ σ₃) &≋ (σ₁ ⊗ σ₂) ⊕ (σ₁ ⊗ σ₃) \tag{$\DISTR$}\\
    \Lower{α₁} ⊗ \Lower{α₂} &≋ \Lower*{α₁ \times α₂} \tag{$\LOWER_⊗$}\\
    \Lower{α₁} ⊕ \Lower{α₂} &≋ \Lower*{α₁ + α₂} \tag{$\LOWER_⊕$}\\
    \Lower{()} ⊗ σ &≋ σ \tag{$\LUNIT_⊗$}\\
    \Lower{\Void} ⊕ σ &≋ σ \tag{$\LUNIT_⊕$} \\
    \Lower{\Void} ⊗ σ &≋ \Lower{\Void} \tag{$\LZERO$}
\end{align*}
\caption{Inductive presentation of open type equivalence. The relation $σ ≋ τ$
  is an \emph{Abelian rig}---a ring without negation---where $\LOWER$ is
  a map from finite types to open quantum types that respects both addition and
  multiplication. }
\label{fig:inductive-equiv}
\end{figure}

\begin{lemma}
  If $f : σ ≋ τ$ and $b : [σ]^\BVar$, then $γ_τ^m(\hat{f} b) = γ_σ^m(b)$.
\end{lemma}
\begin{proof}
  By induction on $f$.
\end{proof}

To complete the proof of \cref{lem:context-cohesion} we need to show that
$σ ⇆ τ$ implies $σ ≋ τ$. We show:
\begin{enumerate}
\item Every open quantum type $σ$ corresponds to one in a
  normal form $N_σ$ such that $σ ≋ N_σ$.
\item If $N_σ ⇆ N_τ$ then $N_σ ≋ N_τ$.
\end{enumerate}
So, if $σ ⇆ τ$ then by (1) it is the case that $σ ≋ N_σ$ and $τ ≋ N_τ$. This
implies $σ ⇆ N_σ$ and $τ ⇆ N_τ$, and so $N_σ ⇆ σ ⇆ τ ⇆ N_τ$. By (2) we can
conclude $σ ≋ N_σ ≋ N_τ ≋ τ$.

Normal quantum types $N$ have the following structure:
\begin{align*}
    \left(\Lower{α₁} ⊗ X^1₁ ⊗ ⋯ ⊗ X^1_{n_1} \right) ⊕
    \cdots 
    ⊕ \left(\Lower{αₘ} ⊗ X^m_1 ⊗ ⋯ ⊗ X^m_{n_m}\right)
\end{align*}

\begin{proposition}
   For every $σ$, there is a normal quantum type $N_σ$ such that $σ ≋ N_σ$.
\end{proposition}
\begin{proof}
  First we define $N_σ$ by induction on $σ$.
  \[ \begin{aligned}
    N_X &≡ \Lower{()} ⊗ X \\
    N_{\Lower{α}} &≡ \Lower{α} \\
    N_{σ ⊕ τ} &≡ N_σ ⊕ N_τ
  \end{aligned} \qquad \begin{aligned}
    N_{σ ⊗ τ} &≡ ⨁_{1 ≤ i ≤ n,1 ≤ j ≤ m} \Lower*{αᵢ \times βⱼ} ⊗ \vec{X^i} ⊗ \vec{Y^j} \\ 
        &\text{where~} N_σ ≡ ⨁_{1 ≤ i ≤ n} \Lower{αᵢ} ⊗ \vec{X^i} \\
    &\text{and~} N_τ ≡ ⨁_{1 ≤ j ≤ m} \Lower{βⱼ} ⊗ \vec{Y^j}
  \end{aligned} \]
  To complete the proof we check that $N_{σ ⊗ τ}≋ N_σ ⊗ N_τ$, which
  follows from distributivity of $⊗$ over $⊕$.
\end{proof}

Now, let $f : N ⇆ N'$ where 
$
    N ≡ ⨁_{1 ≤ i ≤ n} \left( \Lower{αᵢ} ⊗ \vec{X}ᵢ\right) 
$
and
$
    N' ≡ ⨁_{1 ≤ j ≤ n'} \left( \Lower{βⱼ} ⊗ \vec{Y}ⱼ \right) 
$,
where each $\vec{X}ᵢ$ and $\vec{Y}ⱼ$ are $⊗$-separated sequences of type variables.
That means $f$ has the form
\[
    f : Π_{m : \LTypeVar → \Type}~~
        Σ_{i : ℕ_n} αᵢ \times m(\vec{X}ᵢ)
    ≅   Σ_{j : ℕ_{n'}} βⱼ \times m(\vec{Y}ⱼ).
\]

Let $ℜ_f ⊆ 𝒫(ℕ_n \times ℕ_{n'})$ be a relation defined as follows:
\[
    (i,j) ∈ ℜ_f ~~↔~~ Σ_{a : αᵢ,b : βⱼ} 
    f_{(λ\_.())} (i,a) = (j,b).
\]
That is, $f_{λ\_.()}$ has type $Σi, αᵢ ≅ Σj,βⱼ$ and $(i,j) ∈ ℜ_f$ says
there is some $a : αᵢ$ that $f$ maps to some $b : βⱼ$.

Importantly, this implies a broader property by a parametricity argument:

\begin{proposition} \label{fact:parametricity}
  For any $m₁$ and $m₂$ of type $\LTypeVar → \Type$, and for $a : αᵢ$,
  $x₁ : m₁(\vec{X}ᵢ)$, and $x₂ : m₂(\vec{X}ⱼ)$,
  \begin{align*}
    π₁(f_{m₁}(i,a,x₁)) &= π₁(f_{m₂}(i,a,x₂)) \qquad\text{and}\qquad
    π₂(f_{m₁}(i,a,x₁)) = π₂(f_{m₂}(i,a,x₂)).
  \end{align*}
\end{proposition}
\begin{proof}
  Follows from the abstraction theorem by Uemura~\cite{uemura2017}.
\end{proof}

\begin{lemma} \label{lem:XsYs-equiv}
    If $(i,j) ∈ ℜ_f$ then $\vec{X}_i ≋ \vec{Y}ⱼ$. 
\end{lemma}
\begin{proof}
  First, observe that $\vec{X}ᵢ ⇆ \vec{Y}ᵢ$. For a fixed $m$, let $x : m(\vec{X}ᵢ)$. Now,
  take $a$ to be the element of $αᵢ$ witnessed by $(i,j) ∈ ℜ_f$. Then by
  \cref{fact:parametricity} we know that there exists some (unique)
  $b : βⱼ$ and $y : m(\vec{Y}ⱼ)$ such that $f_m(i,a,x) = (j,b,y)$. The map $x ↦ y$
  is in fact an equivalence.

  It is easy to see, then, that $\vec{X}ᵢ ≋ \vec{Y}ⱼ$, by induction on the sizes of
  $\vec{X}ᵢ$ and $\vec{Y}ⱼ$.
\end{proof}

Finally, we can prove the main property of this section.
\begin{proof}[Proof of \cref{lem:context-cohesion}]
The proof is by induction on $n + n'$. We consider five cases: either $ℜ_f$ is an isomorphism, or it is either not functional, not well-defined on all input, not injective, or not surjective. Since $ℜ_f$ is finite, this property is decidable.

\begin{enumerate}

\item Suppose $ℜ_f$ is an isomorphism. Then, observe that whenever
  $(i,j) ∈ ℜ_f$, we have $αᵢ ≅ βⱼ$. This isomorphism is witnessed by the map
  $a ↦ π₂ (f_{λ\_.()}(i,a))$; since $ℜ_f$ is injective, we can be sure that this
  value is in $βⱼ$. Then, applying this fact as well as \cref{lem:XsYs-equiv} we
  have that
  \begin{align*}
    N &= ⨁_{1 ≤ i ≤ n} \left( \Lower{αᵢ} ⊗ \vec{X}ᵢ\right)
    = ⨁_{1 ≤ i ≤ n} \left( \Lower{β_{ℜ_f(i)}} ⊗ \vec{X}ᵢ\right) \\
    &≋ ⨁_{1 ≤ i ≤ n} \left( \Lower{β_{ℜ_f(i)}} ⊗ \vec{Y}_{ℜ_f(i)}\right) 
    ≋ N'
  \end{align*}

\item Suppose $ℜ_f$ is not functional, meaning that there exits some
  $(i,j₁) ∈ ℜ_f$ and $(i,j₂) ∈ ℜ_f$ with $j₁ ≠ j₂$. We know
  $\vec{Y}_{j₁} ≋ \vec{X}ᵢ ≋ \vec{Y}_{j₂}$ by \cref{lem:XsYs-equiv}, so we have that
  \small
  \begin{align*} 
    N' ≋ &\left(\Lower{β_{j₁}} ⊗ \vec{Y}_{j₁}\right) 
        ⊕ \left(\Lower{β_{j₂}} ⊗ \vec{Y}_{j₂}\right)
        ⊕ ⨁_{j ≠ j₁,j₂} \left( \Lower{β_{j}} ⊗ \vec{Y}_{j} \right) \\
    ≋ &\left(\Lower*{β_{j₁} + β_{j₂}} ⊗ \vec{Y}_{j₁}\right)
     ⊕\!\!⨁_{j ≠ j₁,j₂} \left( \Lower{β_{j}} ⊗ \vec{Y}_{j} \right) 
  \end{align*} \normalsize
  Call this new normal type $N''$. 
  We still have $N'' ⇆ N$, but the number of clauses of $N''$ is smaller than that of $N$, so we can invoke the induction hypothesis to show $N'' ≋ N$. By transitivity, $N ≋ N'$.

\item If $ℜ_f$ is not injective, we invoke a similar argument to the case that
  $ℜ_f$ is not functional by reducing the number of clauses of $N$ instead of
  $N'$.

\item Suppose $ℜ_f$ is not well-defined on its domain, meaning that there is
  some $i₀$ not in the domain of $ℜ_f$. Observe first that $α_{i₀}$ must be
  equal to the empty type, $\Void$. If not, then there is some $a : α_{i₀}$, and
  let $j = π_1(f(i₀,a))$; we have $(i₀,j) ∈ ℜ_f$, a contradiction. Thus
  \begin{align*}
    N &≋ \left(\Lower{α_{i₀}} ⊗ \vec{X}_{i₀} \right) 
        ⊕ ⨁_{i ≠ i₀} \left( \Lower{αᵢ} ⊗ \vec{X}ᵢ\right)
      ≋ \left(\Lower{\Void} ⊗ \vec{X}_{i₀} \right) 
        ⊕ ⨁_{i ≠ i₀} \left( \Lower{αᵢ} ⊗ \vec{X}ᵢ\right) \\
      &≋ \Lower{\Void} ⊕ ⨁_{i ≠ i₀} \left( \Lower{αᵢ} ⊗ \vec{X}ᵢ\right)
      ≋ ⨁_{i ≠ i₀} \left( \Lower{αᵢ} ⊗ \vec{X}ᵢ\right) \\
  \end{align*}
  Again, call this new type $N''$. By the induction hypothesis,
  $N'' ≋ N'$, and by transitivity $N ≋ N'$.

\item If $ℜ_f$ is not surjective, the proof follows parallel to the case that
  $ℜ_f$ is not well-defined.

\end{enumerate}
\end{proof}


\subsection{Axioms of partial initialization and measurement.}

Having established \cref{lem:context-cohesion}, we can finally complete the
equational theory for our quantum language by defining two axioms, written
$∼_q$, about the behavior of partial initialization and partial measurement.

For $f : σ ⇆ τ$ and $m : \LTypeVar → \Type$, let us write
$[f]^m : \point{[σ^m]} = \point{[τ]^m}$ for $\ap*[_\POINT]{\univ{f_m}}$.

\begin{axiom} \label{ax:semantic-axioms} 
  Let $f : σ ⇆ τ$ and $b : [σ]^\BVar$, and let 
\small
$e : \LExp{Γ}*{\point{[σ]^m}}$ and
$bs:\prod_{b' : [τ]^\BVar} \LExp*{Γ',γ_τ^m(b')}*{τ}$.
\normalsize
  \begin{align}
      [f]^m ♯ \init[_σ^m]{b} &∼_q \init*[_τ^m]{f_\BVar b} 
        \tagsc{U-intro}\label{eqn:U-intro} \\
    \matchwith{τ}{[f]^m ♯ e}{bs}
                &∼_q \matchwith{σ}{e}{(bs ∘ f_\BVar)}
        \tagsc{U-elim} \label{eqn:U-elim}
  \end{align}
\end{axiom}

\begin{definition}
  Define the relation $e₁ ≈_q e₂$ on expressions as 
  $
    ≈_q ≡ ∼_α ∪ ∼_β ∪ ∼_η ∪ ∼_{cc} ∪ ∼_q.
  $
  We write $e₁ ≈ e₂$ for equality modulo $≈_q$, \ie the type
  $[e₁]_{≈_q} = [e₂]_{≈_q}$.
\end{definition}

\subsection{Instances of equational axioms}

\begin{proposition}[\labelcref{ax:SWAP-intro}~\text{and}~\labelcref{ax:SWAP-elim}]
  \label{prop:SWAP-intro-elim}
  Let $\SWAP$ be the unitary $[\swap]^m$, where $\swap$ is the
  equivalence $λ(x,y).(y,x)$ of type $X ⊗ Y ⇆ Y ⊗ X$. Then 
  \begin{align*}
    \SWAP ♯ (e₁,e₂) &≈ (e₂,e₁) 
        \tag{\ref{ax:SWAP-intro}}
    \\
    \letin{(y,x)}{\SWAP ♯ e}{e'} &≈ \letin{(x,y)}{e}{e'}
        \tag{\ref{ax:SWAP-elim}}
  \end{align*}
\end{proposition}
\begin{proof}
  For the introduction rule, it suffices to show that $\SWAP ♯ (x,y) ≈ (y,x)$
  for free variables $x$ and~$y$.
  \begin{align*}
    \SWAP ♯ (x,y) &≡ \SWAP ♯ \init[_{X ⊗ Y}]{(x,y)} 
    ∼_q \init*[_{Y ⊗ X}]{\swap{(x,y)}} \tag{\ref{eqn:U-intro}} \\
    &≡ \init[_{Y ⊗ X}]{(y,x)} 
    ≡ (y,x)
  \end{align*}
  Elimination is similarly straightforward from Axiom~\ref{eqn:U-elim}.
  \begin{align*}
    \letin{(y,x)}{\SWAP ♯ e}{e'}
    &≡ \matchwith{Y ⊗ X}{(\SWAP ♯ e)}{λ (y,x).~e'} \\
    &∼_q \matchwith{X ⊗ Y}{e}{\left[(λ (y,x).~e') ∘ \swap\right]} \\
    &≡ \matchwith{X ⊗ Y}{e}{λ (x,y).~ e'} 
    &&≡ \letin{(x,y)}{e}{e'} \qedhere
  \end{align*}
\end{proof}


\begin{proposition}
  Let $\CNOT$ be the unitary $\ap[_\POINT]{\cnot}$, where $\cnot$ is the
  equivalence 
  \[ λ (b,b').~(b,\IfThenElse{b}{¬b'}{b'})\] %
  of type $\Bool \times \Bool ≅ \Bool \times \Bool$. Then:
\begin{align*}
    \CNOT ♯ (\put{b},e) &≈~ (\put{b}, \IfThenElse{b}{X ♯ e}{e}) 
             \tagsc{\CNOT-intro} \label{ax:CNOT-intro} \\
     \letin{(!\_,y)}{\CNOT ♯ e}{e'}
      &≈ \letin{(!b,y')}{e}{\IfThenElse{b}{e'\{X ♯ y'/y\}}{e'\{y'/y\}}}
            \tagsc{\CNOT-elim} \label{ax:CNOT-elim}
\end{align*}
\end{proposition}
\begin{proof}
  Let $\DISTR$ be the unitary $[\distr]^m$ where $\distr : \Lower{\Bool} ⊗ X ⇆ X
  ⊕ X$ is defined by
  \[ λ (b,x).~ \IfThenElse{b}{\inr~x}{\inl~{x}}.\] %
  From \cref{eqn:U-intro} we can derive that for booleans $b$ and
  expressions $e$, we have
  \[ \begin{aligned}
    \DISTR ♯ (\put{b},e) &∼_q \IfThenElse{b}{ι₁~e}{ι₀~e} \\
  \end{aligned} \quad\text{and}\quad \begin{aligned}
    \DISTR^{-1} ♯ (ι₀ e) &∼_q (\put{\false},e) \\
    \DISTR^{-1} ♯ (ι₁ e) &∼_q (\put{\true},e)
  \end{aligned} \]

  As a matrix, $\CNOT$ is equal to
  $\DISTR^{-1} ∘ (I ⊕ X) ∘ \DISTR$. Thus,
  \begin{align*}
    \CNOT ♯ (\put{b},e)
    &= \DISTR^{-1} ♯ (I ⊕ X) ♯ \DISTR ♯ (\put{b},e) 
        \tag{\ref{ax:U-compose}}\\
    &≈ \DISTR^{-1} ♯ (I ⊕ X) ♯ \IfThenElse{b}{ι₁~e}{ι₀~e} 
        \tag{\ref{eqn:U-intro}} \\
    &= \IfThenElse{b}{\left(\DISTR^{-1} ♯ (I ⊕ X) ♯ ι₁~e\right)}
                     {\left(\DISTR^{-1} ♯ (I ⊕ X) ♯ ι₀~e\right)} \\
    &= \IfThen{b}{\left(\DISTR^{-1} ♯ ι₁(X ♯ e)\right)}
                 {\left(\DISTR^{-1} ♯ ι₀(I ♯ e)\right)}
        \tagsc{U-⊕-intro} \\
    &≈ \IfThenElse{b}{(\put{\true},X ♯ e)}{(\put{\false},e)} 
        \tagsc{\ref{eqn:U-intro}} \\
    &= (\put{b},\IfThenElse{b}{X ♯ e}{e})
  \end{align*}
  \normalsize

  There is a similar argument for the elimination form. From
  \cref{eqn:U-elim} we can derive that
  \begin{align*}
     \caseof{\DISTR ♯ e}{ι₀ y → e₀}{ι₁ y → e₁}
    &~∼_q~ \letin{(!b,y)}{e}{\IfThenElse{b}{e₁}{e₀}} \\
    \letin{(x,y)}{\DISTR^{-1} ♯ e}{x ⸖ f}
    &~∼_q~ \caseof{e}{ι₀ y → f (\false)}{ι₁ y → f (\true)}
  \end{align*}
  Then:
  \begin{align*}
    \letin{(!\_,y)&}{\CNOT ♯ e}{e'} 
    = \letin{(!\_,y)}{\DISTR^{-1} ♯ I ⊕ X ♯ \DISTR ♯ e}{e'} 
        && \text{(\ref{ax:U-compose})} \\
    &≈ \caseof*{I ⊕ X ♯ \DISTR ♯ e}{ι₀ y → e'}{ι₁ y → e'} 
        && (\text{\ref{eqn:U-elim}}) \\
    &= \caseof*{\DISTR ♯ e}{ι₀ y' → e'\{I ♯ y' / y\}}{ι₁ y' → e'\{X ♯ y' / y\}} 
        && (\text{\ref{eqn:U-oplus-elim}})\\
    &≈ \letin{(!b,y)}{e}{\IfThenElse{b}{e'\{X ♯ y' / y\}}{e'\{y'/y\}}}
        && (\text{\ref{eqn:U-elim}}) \qedhere
  \end{align*}
\end{proof}


\section{Soundness}
\label{sec:semantics}
In this section we give a denotational semantics for the quantum term calculus
with respect to superoperators over density matrices~\citep[Chapter
2]{nielsen2010}. A density matrix is a real-valued square matrix whose trace
sums to one; every density matrix has the form
$ ρ ≡ \sum_j p_j \ket{φ_j}\bra{φ_j},$ where each $\ket{φ_j}$ is a pure state
vector and the coefficients $p_j$ sum to one. A superoperator is a completely
positive map over density matrices that does not increase the trace of its
input.

For a quantum type $σ : \LType$, we define the type $\Density{σ}$ of density
matrices of type $σ$ by quotient induction. First, define
$\Density*{\point{α}}$ to be the collection of density matrices of type
$\Matrix(α,α)$. Next, we must show that for any unitary
$U : \UMatrix(α,β)$, we have
$\Density*{\point{α}}=\Density*{\point{β}}$. This path can be obtained through
univalence from the equivalence $U^∗ ≡ λ ρ. U ρ U^\dagger$. The function $U^∗$ is a
superoperator when $U$ is unitary, and it is invertible via the function
$(U^\dagger)^∗$.

A reader might be concerned that because we defined $\Density{τ}$ by quotient induction, we have inadvertently collapsed all two density matrices $ρ_1,ρ_2$ such that $ρ_2 = U^∗ ρ_1$. This is not true. Just because $ρ : \Density{σ}$ and $\Density{σ}=\Density{σ}$ does not mean that there is a path $ρ = U ρ U^\dagger$. In particular, notice that $\Density{\Qubit}$ is just a set of $2 \times 2$ matrices.

The definition of density matrices can be extended to typing contexts by interpreting a context $Γ ≡ x₁:σ₁,…,xₙ:σₙ$ as a quantum type $σ₁ ⊗ ⋯ ⊗ σₙ$. That is, we write $\Density{Γ}$ for $\Density*{σ₁ ⊗ ⋯ ⊗ σₙ}$. 

The category of superoperators over density matrices is dagger compact
closed~\citep{selinger2007}, has sums (given by the direct product), and has a
terminal object ($\Density*{\Lower{()}}$). We sketch some of its
properties:
\begin{enumerate}[-]

\item For superoperators $f,g : \Density{σ} → \Density{τ}$, pointwise addition $f + g$ is $(f+g)ρ ≡ fρ + gρ$.

\item Superoperators are strict symmetric monoidal, which means that for 
  $f : \Density{σ} → \Density{τ}$ and $g : \Density{σ'} → \Density{τ'}$, we can
  define $f ⊗ f' : \Density*{σ ⊗ σ'} → \Density*{τ ⊗ τ'}$. The unit of $⊗$ is
  $\Density{\Lower{()}}$.  The strictness condition means that
  $\Density*{σ₁ ⊗ (σ₂ ⊗ σ₃)}$ is equal to $\Density*{(σ₁ ⊗ σ₂) ⊗ σ₃}$,
  and that $\Density*{\Lower{()} ⊗ σ}$ is equal to $\Density{σ}$.

\item $\Density*{\Lower{()}}$ is a terminal element, meaning that for every $σ$ there
  is a unique map $! : \Density{σ} → \Density*{\Lower*{}}$ that takes the trace of
  its input matrix.

\item For every $σ₁$ and $σ₂$ there are maps
  $ιᵢ^∗ : \Density{σᵢ} → \Density*{σ₁ ⊕ σ₂}$ defined as $λ ρ. ιᵢ ρ ιᵢ^\dagger$, where
  $ι₁ ≡ \left(\begin{smallmatrix} 1 & 0 \\ 0 & 0 \end{smallmatrix} \right)$ and
  $ι₂ ≡ \left(\begin{smallmatrix} 0 & 0 \\ 0 & 1 \end{smallmatrix} \right)$
  respectively. Given $f₁ : \Density{σ₁} → \Density{τ}$ and
  $f₂ : \Density{σ₂} → \Density{τ}$, there is a unique function
  $[f₁,f₂] : \Density*{σ₁ ⊕ σ₂} → \Density{τ}$ that commutes with $ιᵢ^∗$ in the
  usual way, defined as
  $
    [f₁,f₂] ρ ≡ ι₁^∗ ρ + ι₂^∗ ρ.
  $

\item Generalizing the notion of sum, if $aᵢ$ is the $i$th element of a finite
  type $α$, there is a matrix $δ_{aᵢ} : \Density*{\point{α}}$ characterized
  by a value $1$ at index $(i,i)$ and $0$ elsewhere. For any function $f : α →
  \Density{τ}$, there is a map $[f] : \Density*{\point{α}} → \Density{τ}$
  defined by
  $
    [f] ρ ≡ \sum_{a : α} δ_{a}^∗ ρ.
  $

\end{enumerate}

\subsection{Semantics of quantum expressions}

The denotational semantics of the quantum term language maps expressions $e :
\LExp{Γ}{σ}$ to a superoperator $⟦e⟧ : \Density{Γ} → \Density{q}$ between
density matrices. We define $⟦e⟧$ by induction on $e$.


\begin{enumerate}[-]

\item 
For a variable $x:σ ⊢ x : σ$,  $⟦x⟧$ is the identity function.

\item 
For a let binding $Γ,Γ' ⊢ \letin{x}{e}{e'} : τ$ where $Γ ⊢ e : σ$ and $x:σ,Γ' ⊢
e' : τ$, define $⟦\letin{x}{e}{e'}⟧$ as $⟦e'⟧ ∘ (⟦e⟧ ⊗ \ID)$.


\item
The category of density matrices is symmetric monoidal, which tells us how to
interpret the multiplicative unit and product. That is, given $Γ₁ ⊢ e₁ : σ₁$ and
$Γ₂ ⊢ e₂ : σ₂$, define $⟦Γ₁,Γ₂ ⊢ (e₁,e₂) : σ₁ ⊗ σ₂⟧$ as $⟦e₁⟧ ⊗ ⟦e₂⟧$.

\item
Given $Γ,Γ' ⊢ \letin{(x₁,x₂)}{e}{e'} : τ$ where $Γ ⊢ e : σ₁ ⊗ σ₂$
and $x₁:σ₁,x₂:σ₂,Γ' ⊢ e': τ$,  define $⟦\letin{(x₁,x₂)}{e}{e'}⟧$ as
$⟦e'⟧ ∘ (⟦e⟧ ⊗ \ID)$.




\item 
Given $Γ ⊢ e : σᵢ$, define $⟦Γ ⊢ ιᵢ e : σ₁ ⊕ σ₂⟧$ as $\tilde{ιᵢ} ⟦e⟧$.

\item
Given $Γ ⊢ e : σ₁ ⊕ σ₂$ and $xᵢ:σᵢ,Γ' ⊢ eᵢ : τ$, define
$⟦Γ,Γ' ⊢ \caseof{e}{ι₁ x₁ → e₁}{ι₂ x₂ → e₂} : τ⟧$ as
\[[⟦e₂⟧,⟦e₁⟧] ∘ \DISTR^∗ ∘ (⟦e⟧ ⊗ \ID),\] where
$\DISTR : \UMatrix((α₁ + α₂) \times β, (α₁\timesβ) + (α₂ \times β))$.


\item 
Define $⟦∅ ⊢ \put{a} : \Lower{α}⟧$ as $λ \_.δ_a$. 

\item
Given $Γ ⊢ e : \Lower{α}$ and $f : α → Γ' ⊢ - : τ$, define $⟦Γ,Γ' ⊢ e ⸖ f : τ⟧$
as $\sum_{a : α} ⟦f a⟧ ∘ (⟦e⟧ ⊗ \ID)$.

\end{enumerate}

\subsection{Soundness of the equational theory}

The soundness of $β$, $η$, and commuting conversion equivalences comes down to
the fact that the category of density matrices is symmetric monoidal, has sums,
and has a terminal object.

\begin{theorem}
  If $e₁ ∼_{o} e₂$ for $o ∈ \{β,η,cc\}$, then $⟦e₁⟧ = ⟦e₂⟧$.
\end{theorem}

For any $U : σ = τ$, define $⟦U⟧ : \Density{σ} → \Density{τ}$ trivially by path induction such that $⟦1⟧ = λ x.x$.

\begin{lemma}
    If $U : σ = τ$ and $Γ ⊢ e : σ$, then $⟦U ♯ e⟧ = ⟦U⟧ ∘ ⟦e⟧$.
\end{lemma}

We can now prove the soundness of the axioms regarding the behavior of
unitary equivalences

\begin{restatable}[Soundness of \cref{ax:semantic-axioms}]{theorem}{soundness} \label{thm:axiom-soundness}
  Let $f : σ ⇆ τ$ and $b : [σ]^\BVar$. Then
  \[
    ⟦\widetilde{f_m} ♯ \init[_σ^m]{b}⟧ = ⟦\init*[_τ^m]{f_\BVar b}⟧.
  \]
  If $e : \LExp{Δ}*{\point{[σ]^m}}$ and
  $bs : \prod_{bs : [τ]^\BVar} \LExp*{γ_τ^m(b),Δ'}{q}$, then
  \[ 
    ⟦\matchwith{τ}{(\widetilde{f_m} ♯ e)}{bs}⟧
    = ⟦\matchwith{σ}{e}{bs ∘ f_\BVar}⟧.
  \]
\end{restatable}
\begin{proof}
    
  \cref{sec:normal-sigmas} introduces an inductively-defined relation $σ ≋ τ$
  that holds exactly when $σ ∼ τ$. Thus, it suffices to prove this property with
  respect to $f : σ ≋ τ$. First we check the properties with respect to
  reflexivity, symmetry, transitivity, and congruence. 

  Reflexivity and symmetry follow directly from \cref{prop:U-I} and
  \cref{prop:U-compose} respectively, and congruence follows from the congruence
  of density matrices.


  Next we check the behavior of the ten specific unitaries.
  \cref{fig:semantic-axioms-meas} lists the equations we expect to hold; these
  equalities follow from the properties of density matrices and by unfolding
  definitions. \qedhere

\begin{figure*}
    \[ \begin{aligned}
        ⟦ \SWAP_⊗ ♯ (e₁,e₂) ⟧ &= ⟦(e₂,e₁)⟧ \\
        ⟦ \SWAP_⊕ ♯ ιᵢ e⟧ &= ⟦ ι_{¬i} e⟧ \\
        ⟦ \ASSOC_⊗ ♯ (e₁,(e₂,e₃))⟧ &= ⟦((e₁,e₂),e₃)⟧ \\
        ⟦ \ASSOC_⊕ ♯ ι₀ e ⟧ &= ⟦ι₀ (ι₀ e)⟧ \\
        ⟦ \ASSOC_⊕ ♯ ι₁ (ι₀ e) ⟧ &= ⟦ι₀ (ι₁ e)⟧ \\
        ⟦ \ASSOC_⊕ ♯ ι₁ (ι₁ e) ⟧ &= ⟦ι₁ e⟧ \\
    \end{aligned} \qquad \begin{aligned}
        ⟦ \DISTR ♯ (e₁,ιᵢ e₂)⟧ &= ⟦ιᵢ(e₁,e₂)⟧  \\
        ⟦ \LOWER_⊗ ♯ (\put{a₁},\put{a₂})⟧ &= ⟦\put{(a₁,a₂)}⟧ \\
        ⟦ \LOWER_⊕ ♯ ιᵢ(\put{a})⟧ &= ⟦ \put[~]{\inj{i}{a}} ⟧ \\
        ⟦ \LUNIT_⊗ ♯ (\put{()},e)⟧ &= ⟦e⟧ \\
        ⟦ \LUNIT_⊕ (ι₀ (\put[~]{a : \Void})) ⟧ 
            &= ⟦ \init*[_σ^m]{\emptycase{a}} ⟧ \\
        ⟦ \LUNIT_⊕ (ι₁ e) ⟧ &= ⟦e⟧  \\
        ⟦ \LZERO ♯ (\put[~]{a : \Void},e) ⟧ &= ⟦ \put{a} ⟧
    \end{aligned} \]
     \begin{align*}
        ⟦ \letin{(x₂,x₁)}{\SWAP_⊗ ♯ e}{e'} ⟧ 
            &= ⟦ \letin{(x₁,x₂)}{e}{e'} ⟧ \\
        ⟦ \caseof*{\SWAP_⊕ ♯ e}{ι₀ x₀ → e₀}{ι₁ x₁ → e₁} ⟧ 
            &= ⟦ \caseof{e}{ι₀ x₁ → e₁}{ι₁ x₀ → e₀} ⟧ \\
        ⟦ \letin{((x₁,x₂),x₃)}*{\ASSOC_⊗ ♯ e}{e'} ⟧  
            &= ⟦ \letin{(x₁,(x₂,x₃))}{e}{e'} ⟧ \\
        ⟦ \matchwith{X₁ ⊕ (X₂ ⊕ X₃)}*{\ASSOC_⊕ ♯ e}{bs} ⟧ 
            &= ⟦ \matchwith{(X₁ ⊕ X₂) ⊕ X₃}{e}{bs ∘ \ASSOC_⊕} ⟧ \\ 
        ⟦ \letin{!(a,b)}*{\LOWER_⊗ ♯ e}{e'} ⟧ 
            &= ⟦ \letin{(!a,!b)}{e}{e'} ⟧ \\ 
        ⟦ \caseof*{\DISTR ♯ e}{ι₀ (x,y₀) → e₀}{ι₁ (x,y₁) → e₁} ⟧ 
            &= ⟦ \letin{(x,y)}{e}{\caseof{y}{ι₀ y₀ → e₀}{ι₁ y₁ → e₁}} ⟧ \\
        ⟦ \LOWER_⊕ ♯ e ⸖ f ⟧ 
            &= ⟦ \caseof{e}{ι₀ x₀ → x₀ ⸖ f ∘ \inl}{ι₁ x₁ → x₁ ⸖ f ∘ \inr} ⟧ \\
        ⟦ \letin{(!(),x)}{\LUNIT_⊗ ♯ e}{e'} ⟧ 
            &= ⟦ \letin{x}{e}{e'} ⟧ \\
        ⟦ \LUNIT_⊕ e ⟧ 
            &= ⟦ \caseof{e}{ι₀ !(a : \Void) → \_}{ι₁ x → x} ⟧ \\
        ⟦ \LZERO ♯ e ⸖ f ⟧ 
            &= ⟦ \letin{(!a,\_)}{e}{f a} ⟧ 
    \end{align*}
  \caption{Proof that  \cref{ax:semantic-axioms} is sound with respect to initialization and measurement. Note that the equations containing values $a : \Void$ are vacuously
    true. }
  \label{fig:semantic-axioms-meas}
\end{figure*}

\end{proof}


\begin{theorem}[Soundness] \label{thm:soundness}
  If $e₁ ≈ e₂$, then $⟦e₁⟧ = ⟦e₂⟧$.
\end{theorem}


\section{Completeness}
\label{sec:completeness}
This section will draw a formal connection between the HoTT calculus defined in this paper and Staton's algebraic presentation, which is shown in \cref{fig:staton}.
 
The algebric calculus is presented in continuation-passing style, with judgment $Γ ∣ Δ ⊢ t$ where $Γ$ and $Δ$ are linear contexts in the sense of \cref{sec:lambda}. The intuition is that $Γ$ contains continuation variables, to which quantum variables in $Δ$ can be passed. In Staton's original presentation, variables could hold only qubits, but here we allow variables to hold arbitrary tuples. However, we do restrict types that occur in the algebraic calculus to those of the form $\Qubit^n$. We say that a quantum type is \emph{binary} if it either has the form $\Lower{\Bool}$, or has the form $σ_1 ⊗ σ_2$ for binary types $σ_1$ and $σ_2$.

\begin{figure}
\[
  \begin{array}{c}
    \inferrule*[Right=var]
    {~}
    {Γ,x:σ_1⊗⋯⊗σ_n,Γ' ∣ a_1:σ_1,…,a_n:σ_n ⊢ x(a_1,…,a_n)}
  \qquad \qquad
    \inferrule*[Right=⊗]
    {Γ ∣ Δ,a_1:σ_1,a_2:σ_2 ⊢ t}
    {Γ ∣ Δ,a:σ_1⊗σ_2 ⊢ a(a_1,a_2).t}
  \\ \\
    \inferrule*[Right=new]
    {Γ ∣ Δ,a:\Qubit ⊢ t}
    {Γ ∣ Δ ⊢ \new*{a.t} }
  \qquad \qquad
    \inferrule*[Right=meas]
    {Γ ∣ Δ ⊢ t \\ Γ ∣ Δ ⊢ u}
    {Γ ∣ Δ,a:\Qubit ⊢ \meas*{a,t,u}}
  \\ \\
    \inferrule*[Right=U]
    {U : 𝒰(σ,τ) \\ Γ ∣ Δ,b:τ ⊢ t}
    {Γ ∣ Δ,a:σ ⊢ U(a, b.t)}
  \qquad \qquad
    \inferrule*[Right=subst]
    { x:σ ∣ Δ ⊢ t \\ Γ ∣ Θ,a:σ ⊢ u}
    { Γ ∣ Δ,Θ ⊢ t[x(a) ↦ u]}
\end{array}\]
\caption{Algebraic structure of quantum computation.}
\label{fig:staton}
\end{figure}

Initialization, written $\init*{a.t}$, produces a qubit $\ket{0}$, binds it to $a$, and continues as $t$. Measurement $\meas*{a,t,u}$ measures the qubit $a$, and either continues as $t$ if the result is $0$, or as $u$ if the result is $1$.

We choose to encode unitary application using $\TRANSPORT$ as in \cref{sec:types}: $U(a,b.t) ≡ \transport{}{U}{t}\{a/b\}$.
In addition, the substitution rule \textsc{subst} is a derived rule defined by induction on $t$:
\[ \begin{aligned}
    x(a_1,…,a_n)[x(a) ↦ u] &≡ a(a_1,…,a_n).u \\
    y(a_1,…,a_n)[x(a) ↦ u] &≡ y(a_1,…,a_n) \\
    (b(b_1,b_2).t)[x(a) ↦ u] &≡ b(b_1,b_2).t[x(a) ↦ u]
  \end{aligned} \qquad \begin{aligned}
    \new*{b.t}[x(a) ↦ u] &≡ \new*{b.t[x(a) ↦ u]} \\
    \meas*{b,t_0,t_1}[x(a) ↦ u] &≡ \meas*{b,t_0[x(a)↦u],t_1[x(a)↦u]}
\end{aligned}\]

The equational theory for this algebra, which we write $t ≂ u$, is given is
\cref{fig:staton-administrative,fig:staton-interesting}. The first set of
``interesting'' axioms correspond closely to those shown in
\cref{fig:semantic-axioms}. For example, \cref{ax:A} says that measuring a qubit
$X ♯ e$ is the same as measuring $e$ and taking the opposite branch. \cref{fig:staton-interesting} uses the following shorthand:
\[ \begin{aligned}
    \discardin{a}{t} &≡ \meas*{a,t,t} \\
    U(a,t) &≡ U(a,a.t)
  \end{aligned} \qquad\qquad \begin{aligned}
    D(U,V) &: 𝒰(\Qubit ⊗ σ,\Qubit ⊗ σ) \\
        &≡ \DISTR^{-1} ∘ (U ⊕ V) ∘ \DISTR
\end{aligned} \]
The second set of ``administrative'' axioms
correspond to \cref{fig:structural-axioms,fig:groupoid-axioms}.

\begin{figure}
\begin{minipage}{0.45\textwidth}
\vspace{-5mm}
\begin{align*}
    &X(a,\meas*{a,x,y}) ≂ \meas*{a,y,x}
        \label{ax:A}\tag{A} \\
    &\meas*{a,U(b,x(b)),V(b,y(b))} \\
    &≂ D(U,V)((a,b), \meas*{a,x(b),y(b)})
        \label{ax:B} \tag{B}
\end{align*}
\end{minipage}
\begin{minipage}{0.6\textwidth}
\begin{align*}
    U(a,\discardin{a}{x}) &≂ \discardin{a}{x}
        \label{ax:C} \tag{C} \\
    \new*{a.\meas*{a,x,y}} &≂ x
        \label{ax:D} \tag{D} \\
    \new*{a.D(U,V)((a,b),x(a,b))} &≂ U(b,\new*{a,x(a,b)})
        \label{ax:E} \tag{E} \\
\end{align*}
\end{minipage}
\caption{Staton's ``interesting'' axioms regarding the behavior of measurement and initialization~\cite{staton2015}.}
\label{fig:staton-interesting}
\end{figure}

\begin{figure}
\begin{minipage}{0.5\textwidth}
\begin{align*}
    \SWAP((a,b),x(a,b)) &≂ x(b,a)
        \label{ax:F} \tag{F} \\
    I(a,x(a)) &≂ x(a)
        \label{ax:G} \tag{G} \\
    (VU)(a,x(a) &≂ V(a,U(a,x(a))) 
        \label{ax:H} \tag{H} \\
    (U ⊗ V)((a,b),x(a,b)) &≂ U(a,V(b,x(a,b))) 
        \label{ax:I} \tag{I}
\end{align*}
\end{minipage}
\quad
\begin{minipage}{0.55\textwidth}
\begin{align*}
    &\meas*{a,\meas*{b,u,v},\meas*{b,x,y}} \\
        &\quad≂ \meas*{b,\meas*{a,u,x},\meas*{a,v,y}}
        \label{ax:J} \tag{J} \\
    &\new*{a.\new*{b.x(a,b)}} ≂ \new*{b.\new*{a.x(a,b)}}
        \label{ax:K} \tag{K} \\
    &\new*{a.\meas*{b,x(a),y(a)}} \\
      &\quad≂\meas*{b,\new*{a.x(a)},\new*{b.y(b)}}
        \label{ax:L} \tag{L} \\
\end{align*}
\end{minipage}
\caption{Staton's ``administrative'' axioms~\citep{staton2015}.}
\label{fig:staton-administrative}
\end{figure}

Finally, we add three commuting conversion rules to our calculus that define the behavior of the $⊗$ rule, which were not explicit in Staton's original presentation.
\begin{align*}
    a(a_1,a_2).b(b_1,b_2).t &≂ b(b_1,b_2).a(a_1,a_2).t \tag{M} \\
    a(a_1,a_2).\new*{b.t} &≂ \new*{b.a(a_1,a_2).t} \tag{N} \\
    a(a_1,a_2).\meas*{b,t_1,t_2} &≂ \meas*{b,a(a_1,a_2).t_1, a(a_1,a_2).t_2} \tag{O}
\end{align*}

\subsection{Algebraic to HoTT calculus translation}

Let $x:σ ∣ Δ ⊢ t$ be a term with exactly one continuation. Then we can define $\fromS{t} : \LExp{Δ}{σ}$ as follows:
\begin{align*}
    \fromS{x(a_1,…,a_n)} &≡ (a_1,…,a_n) \\
    \fromS{\new*{a.t}} &≡ \letin{a}{\init{0}}{\fromS{t}} \\
    \fromS{\meas*{a,t,u}} &≡ \letin{!x}{a}{\IfThenElse{x}{\fromS{u}}{\fromS{v}}}
\end{align*}

\begin{lemma}
    If $U : 𝒰(σ,τ)$ then $\fromS{U(a,b.t)} ≡ \letin{b}{U♯a}{\fromS{t}}$.
\end{lemma}
\begin{proof}
    By path induction on $U :σ=τ$.
\end{proof}

\begin{theorem}\label{thm:fromSsoundness}
    If $t ≂ u$ then $\fromS{t} ≈ \fromS{u}$
\end{theorem}
\begin{proof}
    Straightforward from the proofs in \cref{sec:types,sec:axioms}.
\end{proof}

\subsection{HoTT to algebraic calculus translation}

Since the algebraic calculus can only represent binary types, we omit the term
constructors for sum types $⊕$, and we write $\BExp{Γ}{τ}$ for this restricted
class of binary expressions.

\cref{fig:HoTTtoAlgebra} defines a translation from a binary expression $e : \BExp*{Γ}{τ}$ to $y:τ ∣ Γ ⊢ \toS[y]{e}$.

\begin{figure*}
\small
\[\begin{array}{c}
    \inferrule*
    {~}
    {x:\BExp*{x:σ}{σ}}
  \quad⇒\quad
    \inferrule*
    {~}
    {y:σ ∣ x:σ ⊢ y(x)}
  \\ \\
    \inferrule*
    { e : \BExp{Γ}{σ} \\
      e' : \BExp*{Γ',x:σ}{τ} }
    { \letin{x}{e}{e'} : \BExp*{Γ,Γ'}{τ} }
  \quad⇒\quad
    \inferrule*
    { y:σ ∣ Γ ⊢ \toS[y]{e} \\
      z:τ ∣ Γ',x:σ ⊢ \toS[z]{e'} }
    { z:τ ∣ \toS[y]{e}[y ↦ \toS[z]{e'}] }
  \\ \\
    \inferrule*
    { e₁ : \BExp{Γ₁}{σ₁} \\
      e₂ : \BExp{Γ₂}{σ₂} }
    { (e₁,e₂) : \BExp*{Γ₁,Γ₂}*{σ₁ ⊗ σ₂} }
  \quad⇒\quad
    \inferrule*
    { y_1:σ_1 ∣ Γ_1 ⊢ \toS{e_1} \\
      y_2:σ_2 ∣ Γ_2 ⊢ \toS{e_2} }
    { y:σ_1 ⊗ σ_2 ∣ Γ_1,Γ_2 ⊢ \toS{e_1}[y_1(x_1) ↦ \toS{e_2}[y_2(x_2) ↦ y(x_1,x_2)]] }
  \\ \\
    \inferrule*
    { e : \BExp{Γ}*{σ₁ ⊗ σ₂} \\
      e' : \BExp*{Γ',x₁:σ₁,x₂:σ₂}{τ} }
    { \letin{(x₁,x₂)}{e}{e'} : \BExp*{Γ,Γ'}{τ} }
  \quad⇒\quad
    \inferrule*
    { y : σ_1 ⊗ σ_2 ∣ Γ ⊢ \toS{e} \\
      z : τ ∣ Γ',x_1:σ_1,x_2:σ_2 ⊢ \toS{e'} }
    { z : τ ∣ Γ,Γ' ⊢ \toS{e}[y(x) ↦ x(x_1,x_2). \toS[z]{e'}]}
  \\ \\
    \inferrule*
    { b : \Bool }
    { \put{b} : \BExp{∅}*{\Qubit} }
  \quad⇒\quad
    \inferrule*
    {b = \false}
    { x:\Qubit ∣ ∅ ⊢ \new*{a.x(a)} }
  \quad,\quad
    \inferrule*
    {b = \true}
    { x:\Qubit ∣ ∅ ⊢ \new*{a.X(a,x(a))} }
  \\ \\
    \inferrule*
    { e : \BExp{Γ}*{\Lower{\Bool}} \\
      f : \Bool → \BExp{Γ'}{τ} }
    { e ⸖ f : \BExp*{Γ,Γ'}{τ} }
  \quad⇒\quad 
    \inferrule*
    { x:\Lower{\Bool} ∣ Γ ⊢ \toS{e} \\
      \prod_b y:τ ∣ Γ' ⊢ \toS{f b} }
    { y:τ ∣ Γ,Γ' ⊢ \toS{e}[x(q) ↦ \meas*{q,f(\false),f(\true)}]}
\end{array}\]
\normalsize
\caption{Encoding of the HoTT quantum calculus defined in this paper into Staton's algebraic calculus.}
\label{fig:HoTTtoAlgebra}
\end{figure*}

\begin{lemma}
  If $e : \BExp{Γ}{σ}$ and $U : σ = τ$ then 
    $\toS[y]{U ♯ e} ≂ \toS[x]{e}{x(a) ↦ U(a,y(a))}.$
\end{lemma}
\begin{proof}
    By path induction on $e$, it suffices to observe that $\toS[y]{e} =
    \toS[x]{e}[x(a) ↦ y(a)]$.
\end{proof}

\begin{theorem}\label{thm:toSsoundness}
  If $e_1,e_2 : \BExp{Γ}{σ}$ and $e_1 ≈ e_2$, then $\toS{e_1} ≂ \toS{e_2}$.
\end{theorem}

\begin{lemma} \label{lem:roundtrip}
    For all $e : \BExp{Γ}{σ}$ and $x:τ ∣ Δ ⊢ t$:
    $\fromS{\toS{e}} ≈ e$ and $\toS[x]{\fromS{t}} ≂ t.$
\end{lemma}

\begin{theorem}
  The $\BEXP$ HoTT calculus is sound and complete with respect to the algebraic calculus.
\end{theorem}
\begin{proof}
    If $e_1 ≈ e_2$ then $\toS{e_1} ≂ \toS{e_2}$ by \cref{thm:soundness}, and if $\toS{e_1} ≂ \toS{e_2}$ then, by \cref{thm:fromSsoundness,lem:roundtrip}, 
  \[ e_1 ≈ \fromS{\toS{e_1}} ≈ \fromS{\toS{e_2}} ≈ e_2. \qedhere \]
\end{proof}

Staton proves that the algebraic calculus is sound and fully complete with respect to ${\CStarCPU}^2$, the category of $\CStar$ algebras of dimension $2^n$ with completely positive and unitary transformations~\citep{staton2015}. As a consequence, the completeness of the $\BEXP$ fragment of the HoTT calculus extends to $\CStarCPU$. We speculate but have yet to prove that the unrestricted HoTT calculus is sound and complete with respect to $\CStarCPU$ algebras of arbitrary dimension.

\section{Discussion}
\label{sec:discussion}
In this section we discuss some of the design decisions made in this work.


\medskip \noindent \textbf{Axiom schemes.} \enspace 
Our equational theory prioritizes equations based on the structure of the
language, such as $β$, $η$, and commuting conversion rules. Such rules do not
depend on any quantum-specific principles, and their meta-theories are
well-understood. In addition, we prioritize collecting many axioms into a single
axiom scheme, as we do for the equational axioms \ref{eqn:U-intro} and
\ref{eqn:U-elim}. This approach gives concise axioms that highlight the
important structure, but requires more overhead to express.


Our axiom schemes are also somewhat redundant---for example, we proved the
equations for the not unitary $X$ in \cref{prop:X-intro-elim}, but they are also
a consequence of the \ref{eqn:U-intro} and \ref{eqn:U-elim} axioms.

\medskip \noindent \textbf{Unitaries.} \enspace 
This work does not axiomatize unitary transformations, in order to focus on the
relationship between quantum and non-quantum data. However, axiomatizations
based on universal (or even non-universal) sets of unitaries, such as those by
Matsumoto and Amano~\cite{matsumoto2008} or Amy et al~\cite{amy2017}, could be incorporated with a
higher-inductive type (HIT). As a first approximation, we could define $\QType$
as a HIT that axiomatizes only the behavior of the Hadamard gate $H$, with the
following constructors: a type $\Qubit : \QType$; a path $\kwfont{H} : (\Qubit =
\Qubit)$, and a higher path expressing that $H^\dagger = H$. Since unitaries are still
encoded in the path type of quantum types, the aspects of the equational theory
we derived by path induction would still hold, and it would allow finer control
over the ways by which unitaries approximate each other. On the other hand,
working with higher inductive types with many constructors can quickly become
unwieldy.

\medskip \noindent \textbf{Quantum types as host types.} \enspace 
It may seem odd to embed arbitrary host language values as quantum data, via the
$\LOWER$ type. In contrast, Quipper has both host-language booleans and embedded
bits, which are different from qubits. We could have designed a similarly
restricted system by removing the $\PUT$ constructor, so that the only way to
construct the $\LOWER$ type is via explicit measurement,
$\MEAS : \Qubit ⊸ \Lower{\Bool}$. However, because we have restricted ourselves
to finite types, all such data is sound with respect to finitary quantum
computing.


\medskip \noindent \textbf{Other equational theories of quantum computing.} \enspace 
Departing from the embedded QRAM model,
other models of quantum computation have elegant equational theories, including
algebraic presentations such as the arrow calculus~\citep{vizzotto2013} and
graphical presentations such as the ZX calculus~\cite{backens2015}. The ZX calculus is not a programming language in the usual sense, but is a graphical calculus that makes associativity and symmetry irrelevant. It's equational theory has been proven sound and complete, but it lacks programming abstractions such as modularity and polymorphism.

On the other hand, the arrow calculus is a $\lambda$-calculus that builds on a body of work allowing programmers to construct density matrices and superoperators directly in a programmatic way, for example by adding matrices together directly~\citep{altenkirch2007,vizzotto2009,vizzotto2009a}. This differs from the style of language presented in this paper, where the only way that users can obtain non-unitary results is explicitly through measurement, which corresponds more closely to the QRAM style of interaction with a quantum computer.

\medskip \noindent \textbf{Conclusion.} \enspace 
This paper presents an equational theory for a linear quantum term calculus in a
compact and elegant style using homotopy type theory. We justify these claims by
deriving an equational theory known to be complete for a less expressive
language, and by proving the semantics is sound with respect to a standard model
of quantum computation. In doing so, we have both introduced a new tool to the
study of quantum equational theory, and also demonstrated the application of
homotopy type theory in a programming environment.


\section*{Acknowledgment}
  The authors are grateful to Matthew Weaver, Antoine Voizard, and Sam Staton
  for discussions regarding this work, as well as the anonymous reviewers who
  have provided feedback. This work is supported in part by the AFRL MURI No.
  FA9550-16-1-0082.

\bibliographystyle{eptcs}
\bibliography{bibliography}

\appendix





\end{document}